\documentclass[aps, pra, superscriptaddress, reprint]{revtex4-1}

\usepackage{graphicx}
\usepackage{dcolumn}
\usepackage{bm}
\usepackage{amsfonts}
\usepackage{amsmath}
\usepackage{amssymb}
\usepackage{color}
\usepackage[normalem]{ulem}
\usepackage[percent]{overpic}
\usepackage{natbib}
\usepackage{soul}

\usepackage[colorlinks=true,citecolor=blue]{hyperref}
\hypersetup{colorlinks=true,citecolor=blue,linkcolor=red,urlcolor=blue}

\def\be{\begin{equation}}
\def\ee{\end{equation}}

\begin{document}

\title{Density and pseudo-spin rotons in a bilayer of soft-core bosons}

\author{Fatemeh Pouresmaeeli}
\email{f.pouresmaeeli@iasbs.ac.ir}
\affiliation{Department of Physics, Institute for Advanced Studies in Basic Sciences (IASBS), Zanjan 45137-66731, Iran}

\author{Saeed H. Abedinpour}
\email{abedinpour@iasbs.ac.ir}
	\affiliation{Department of Physics, Institute for Advanced Studies in Basic Sciences (IASBS), Zanjan 45137-66731, Iran}
	\affiliation{Research Center for Basic Sciences \& Modern Technologies (RBST), Institute for Advanced Studies in Basic Sciences (IASBS), Zanjan 45137-66731, Iran}
\author{B. Tanatar}
	\affiliation{Department of Physics, Bilkent University, Bilkent, 06800 Ankara, Turkey}

\begin{abstract}
We study the dynamics of a bilayer system of bosons with repulsive soft-core Rydberg-dressed interactions 
within the mean-field Bogoliubov-de Gennes approximation.
We find roton minima in both symmetric and asymmetric collective density modes of the symmetric bilayer. 
Depending on the density of bosons in each layer and the spacing between two layers, the homogeneous superfluid phase becomes unstable in either (or both) of these two channels, leading to density and pseudo-spin-density wave instabilities in the system.
Breaking the symmetry between two layers, either with a finite counterflow or a density imbalance renormalizes the dispersion of collective modes and makes the system more susceptible to density-wave instability. 
\end{abstract}
\maketitle

\section{Introduction}\label{intro}
Rydberg atoms have attracted remarkable attention in recent years~\cite{browaeys2020,de,browaeys2016}. 
The ability to control Rydberg interaction in these systems provides a platform to realize artificial many-body systems~\cite{labuhn} and quantum simulators~\cite{weimer,scholl}. 
Rydberg atoms have been employed to investigate various problems in quantum information~\cite{Saffman,Jau}, quantum computation~\cite{Morgado}, and nonlinear quantum optics~\cite{Peyrone,firstenberg}.  
It has been demonstrated recently that it is possible to collectively encode information on Rydberg atoms where the qubit is stored as a superposition of Rydberg polariton modes rather than a single atom~\cite{Spong_PRL2021}.

Rydberg states suffer from short lifetimes which makes their use in the experimental study of the atomic 
dynamics very difficult~\cite{Johnson,balewski}. 
To overcome this problem, the Rydberg state could be mixed with the ground state~\cite{Henkel,glaetzle,mayle}
producing Rydberg-dressed atoms with much longer lifetimes.
The interaction between these Rydberg-dressed particles has a soft-core and a van der Waals tail at long distances. 
Many exotic phenomena such as supersolid phase~\cite{cinti2014,henkel2012}, quantum liquid droplets~\cite{Li_PRL2016,seydi2021}, clustering~\cite{teruzzi}, $Z_{\rm 2}$ topological order~\cite{verresen} and roton excitation~\cite{seydi2020,teruzzi,macri} are investigated with Rydberg-dressed atoms. 
Rotating trapped Rydberg-dressed atoms is suggested to provide an exceptional platform to observe the quantum Hall effect with neutral atomic gases~\cite{Burrello_PRR2020}.

The emergence of maxon-roton structure in the excitation spectrum is a characteristic of strongly correlated Bose liquids~\cite{Petter,chomaz,boudjemaa,Henkel,chen}. 
The roton minimum has been theoretically predicted for quasi two dimensional dipolar Bose-Einstein condensates (BECs) within the mean-field approximation~\cite{Santos} and experimentally observed in Bose-Einstein condensates of highly magnetic erbium atoms~\cite{chomaz}.

Path integral Monte Carlo simulation has been employed to obtain the phase diagram of two-dimensional (2D) ultra-cold Rydberg-dressed bosons. It has been shown that various phases, such as superfluid and supersolid phases appear in the phase diagram by varying the density and coupling strength~\cite{cinti2014}. Two different roton minima have been found in the dispersion of Rydberg-dressed BEC with the spin-orbit coupling and the roton instability becomes more accessible~\cite{lyu}.
The excitation spectrum of Rydberg-dressed bosons has been studied based on the hypernetted-chain Euler-Lagrange approximation in which the density-wave instability is predicted for some range of parameters in both three and two dimensions~\cite{seydi2020}.  
The phase diagram of a two-dimensional binary Rydberg-dressed BEC is also investigated in the super-solid regime ~\cite{Hsueh_PRA2013}.

In this paper, we investigate the dynamics of bosons with Rydberg-dressed interaction in a double layer structure within the Bogoliubov-de Gennes (BdG) mean-field approximation at zero temperature. The maxon-roton behavior appears for two collective modes corresponding to the density and pseudo-spin modes of the bosonic bilayer. When the two layers are decoupled, these modes become degenerate. Such degeneracy is lifted through a finite counterflow or a density imbalance between two layers. 
We also find that density-wave instability could occur at either or both of the total density and pseudo-spin channels for some values of the coupling constant and layer spacing.

In the rest of this paper, we calculate the collective modes of a bilayer system and investigate the density-wave instability for the symmetric as well as density imbalanced bilayer in Sec.~\ref{sec:excitation}. The effects of finite counterflow on the mode dispersion and instability are also investigated in this section. 
We summarize our main findings in Sec.~\ref{sec:results}. 
\section{Quasiparticle spectrum}\label{sec:excitation}
\begin{figure}
	\begin{tabular}{cc}
		\includegraphics[width=\linewidth]{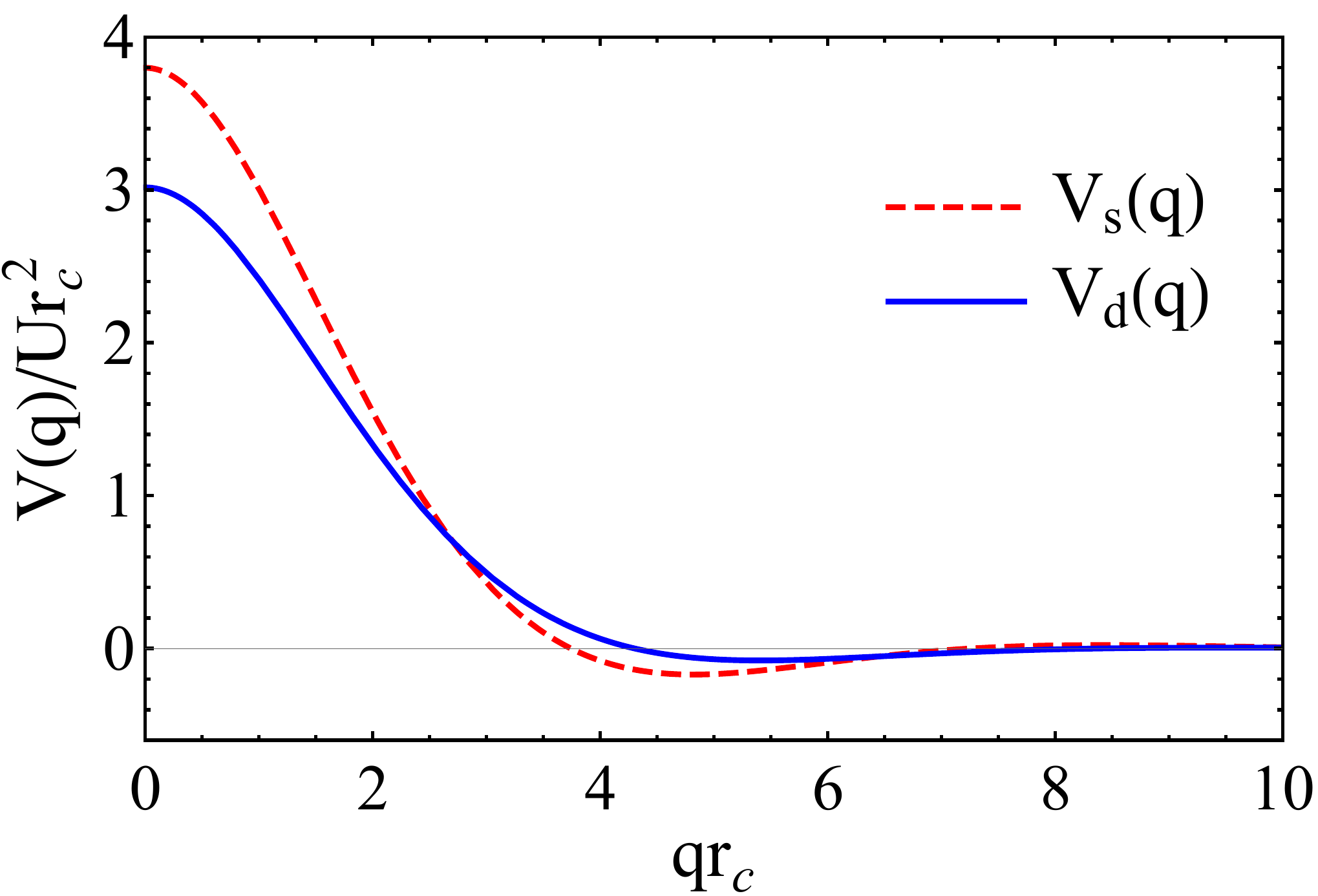}
	\end{tabular}
	\caption{The Fourier transforms of the intralayer and interlayer Rydberg-dressed interactions versus $qr_c$. The layer spacing is $d=0.5\, r_c$ here.
}	\label{fig:Vq_rydberg}
\end{figure}

When we consider bosons with repulsive soft-core Rydberg-dressed interactions loaded in two layers the intralayer and interlayer interactions in the real space read~\cite{khasseh}
\be\label{eq:vr_ryd}
\begin{split}
	V_{\rm s}(r)=& \frac{U}{{1 + {{(r/{r_c})}^6}}}, \\
	V_{\rm d}(r) =& \frac{U}{{1 + {{(\sqrt {{r^2} + {d^2}} /{r_c})}^6}}}.
\end{split}
\ee 
Here, $U$  and ${r_c}$  are the interaction strength and the soft-core radius, respectively, which in turn depend on the Raman coupling, red detuning, and the van der Waals coefficients~\cite{Henkel,Johnson,Honer}, and $d$ is the separation between two layers.
The Fourier transform of $V_{\rm s}(r)$ is a complicated Meijer-G function~\cite{khasseh,seydi2020}, while we have to perform the Fourier transform of  $V_{\rm d}(r)$ numerically. Fig.\,\ref{fig:Vq_rydberg} illustrates the wavevector dependance of $V_s(q)$ and $V_d(q)$.  
The quasiparticle excitation spectrum of a two-component Bose gas in the Bogoliubov-de Gennes (BdG) 
approximation reads~\cite{Pitaevskii} 
\be\label{eq:w_pm}
\hbar \omega_{\pm}(q)=\sqrt{\varepsilon^2_q+2n \varepsilon_q V_\pm(q)},
\ee
where $\varepsilon_q=\hbar^2 q^2/(2 m)$ is the dispersion of free bosons, $n$ is the particle density in each layer (assumed to be the same), and $V_\pm(q)=V_{\rm s}(q)\pm V_{\rm d}(q)$ are the symmetric and asymmetric components of the interaction. The symmetric and asymmetric modes are the total density and pseudo-spin modes, respectively, if we regard the two layers as the pseudo-spin degree of freedom. 
Note that the BdG excitation spectrum is identical to the dispersion of collective modes within the random-phase approximation (see, e.g. Ref.~\cite{abedinpour_LTP2020} for details).
For a two-dimensional Bose gas with Rydberg-dressed interactions, the many-body corrections beyond the mean-field approximation on the excitation spectrum and density-wave instability are mainly quantitative~\cite{seydi2020} and we surmise the same remains true also for a bilayer system. 

Within the mean-field approximation, we can describe our symmetric bilayer system with two parameters i.e., the dimensionless coupling constant $\alpha=nm r_c^4 U/\hbar^2$ and the dimensionless layer spacing ${\tilde d}=d/r_c$.

\subsection{Long wavelength excitations and sound velocity}
\begin{figure}
		\includegraphics[width=\linewidth]{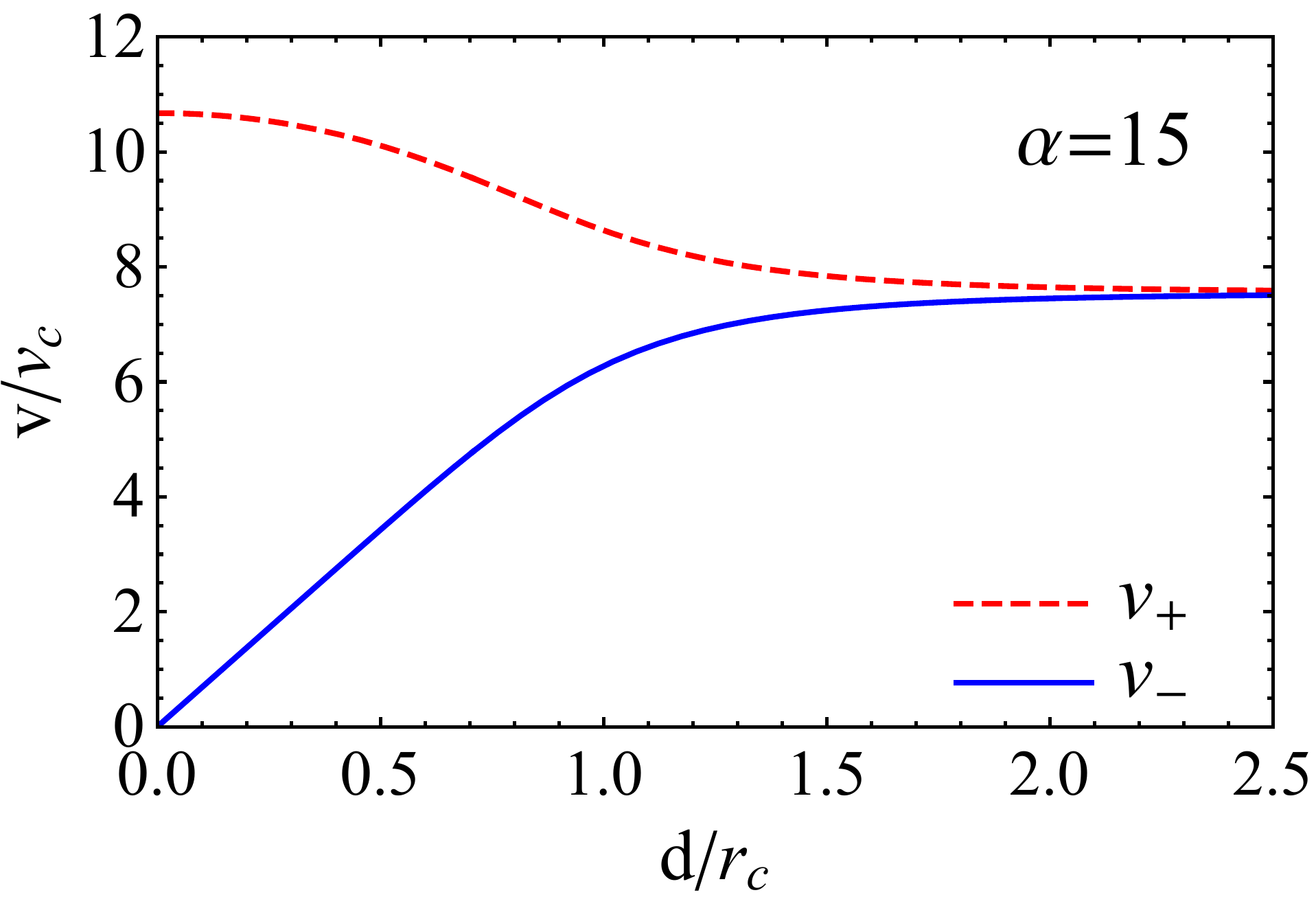}
	\caption{The sound velocities of symmetric and asymmetric modes [in the units of $v_c=\hbar/(mr_c)$] versus $d/r_c$ for $\alpha=15$.
}	\label{fig:sound_velocity_rydberg}
\end{figure}

It follows from Eq.\,(2) that in the long wavelength limit, the dispersion of both symmetric and 
asymmetric modes are linear
\be\label{eq:w_pm_q0}
 \omega_{\pm}(q\to 0)=v_\pm q +{\cal O}(q^2),
\ee
where $v_\pm=\sqrt{nV_\pm(q=0)/m}$ are the sound velocity of the two modes.
From the Fourier transform of the Rydberg-dressed potentials, we find $V_s(q=0)=2\pi^2 U r_c^2/(3 \sqrt{3})$ and
\be
\begin{split}
V_{\rm d}(0)&=\frac{\pi U r_c^2}{6}
\left[\sqrt{3}\pi -2 \sqrt{3}\tan^{-1}\left(\frac{2 \tilde d^2-1}{\sqrt{3}}\right)\right.\\
&\qquad~~~~~~~~~~~~~~~~~~\left.+\ln\left(\frac{1-\tilde d^2+\tilde d^4}{(1+{\tilde d}^2)^2}\right)\right]\\
&\approx V_{\rm s}(0)-\pi U d^2+{\cal O}(d^8).
\end{split}
\ee
At small layer spacings, to linear order in $d$, we find
\be
\begin{split}
v_+(d\to 0)&\approx v_c 2 \pi \sqrt{\frac{\alpha}{3 \sqrt{3}}},\\
v_-(d\to 0)& \approx v_c \sqrt{\pi \alpha} {\tilde d},
\end{split}
\ee
where $v_c=\hbar/(mr_c)$. 
In Fig.\,\ref{fig:sound_velocity_rydberg} we show the layer separation dependance of the sound velocities 
where the $d\rightarrow 0$ behavior is evident. For large $d$ the two modes become degenerate and the sound velocities approach to a common value.

\subsection{Density and pseudo-spin rotons}
While the dispersion of both quasiparticle excitation branches are linear at long wavelengths and free-particle-like 
(i.e., quadratic) at large wave vectors, at intermediate values of $q$ it is possible to observe maxon-roton features 
at strong couplings \cite{Henkel,McCormack}. 
For Rydberg-dressed interaction, these features are already observable in the simple BdG approximation.

\begin{figure}
	\begin{tabular}{cc}
		\includegraphics[width=0.5\linewidth]{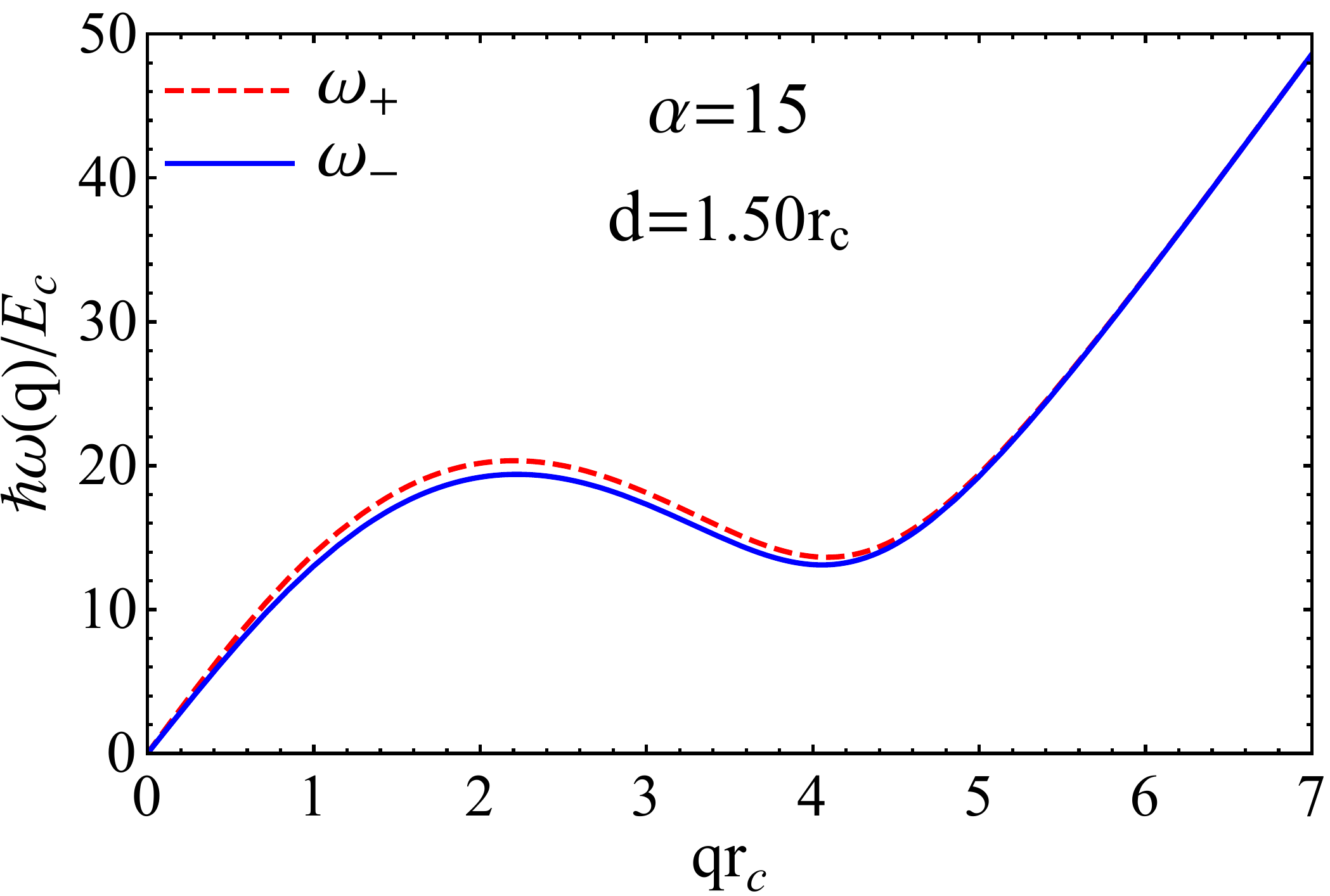}&
		\includegraphics[width=0.5\linewidth]{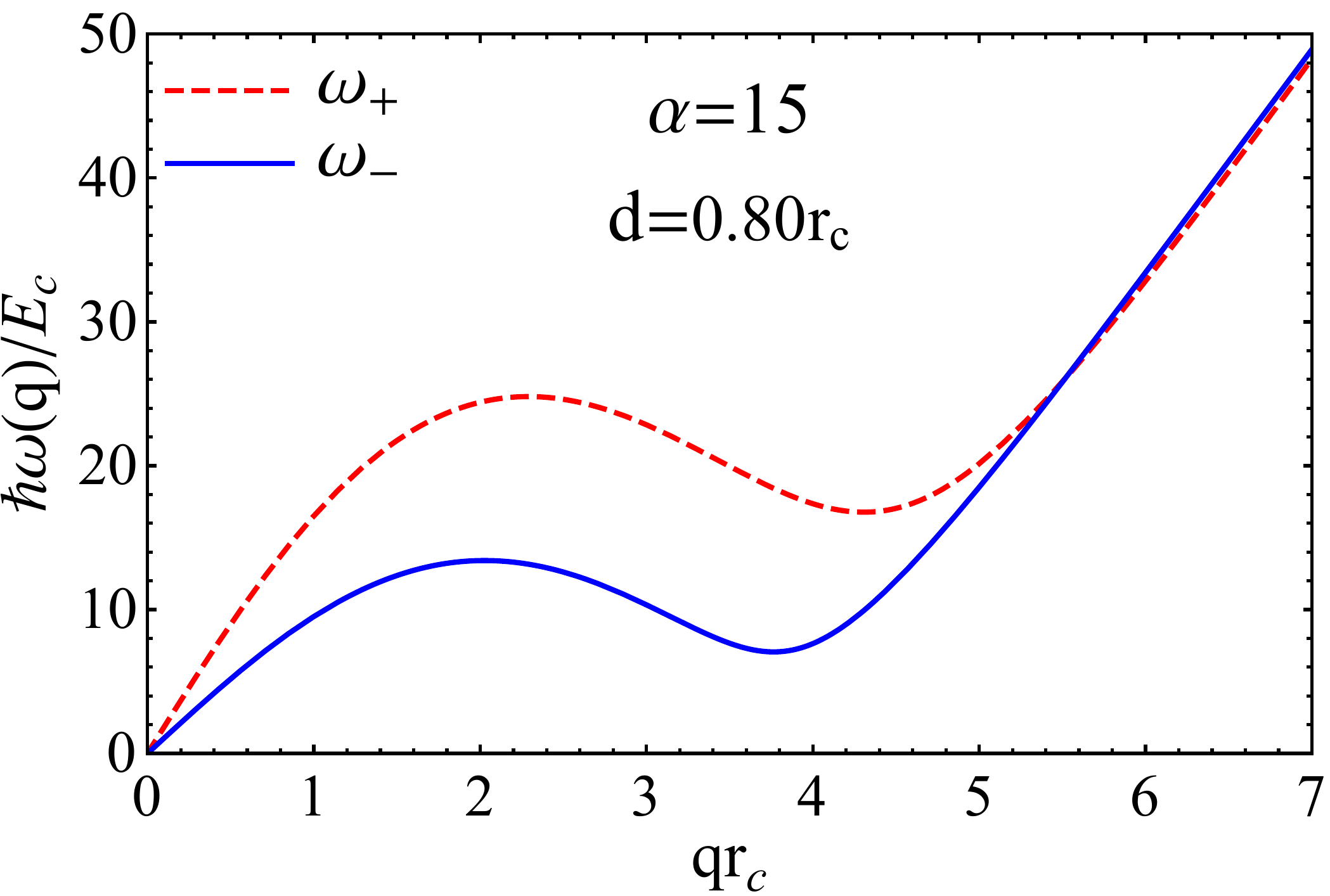}\\
		\includegraphics[width=0.5\linewidth]{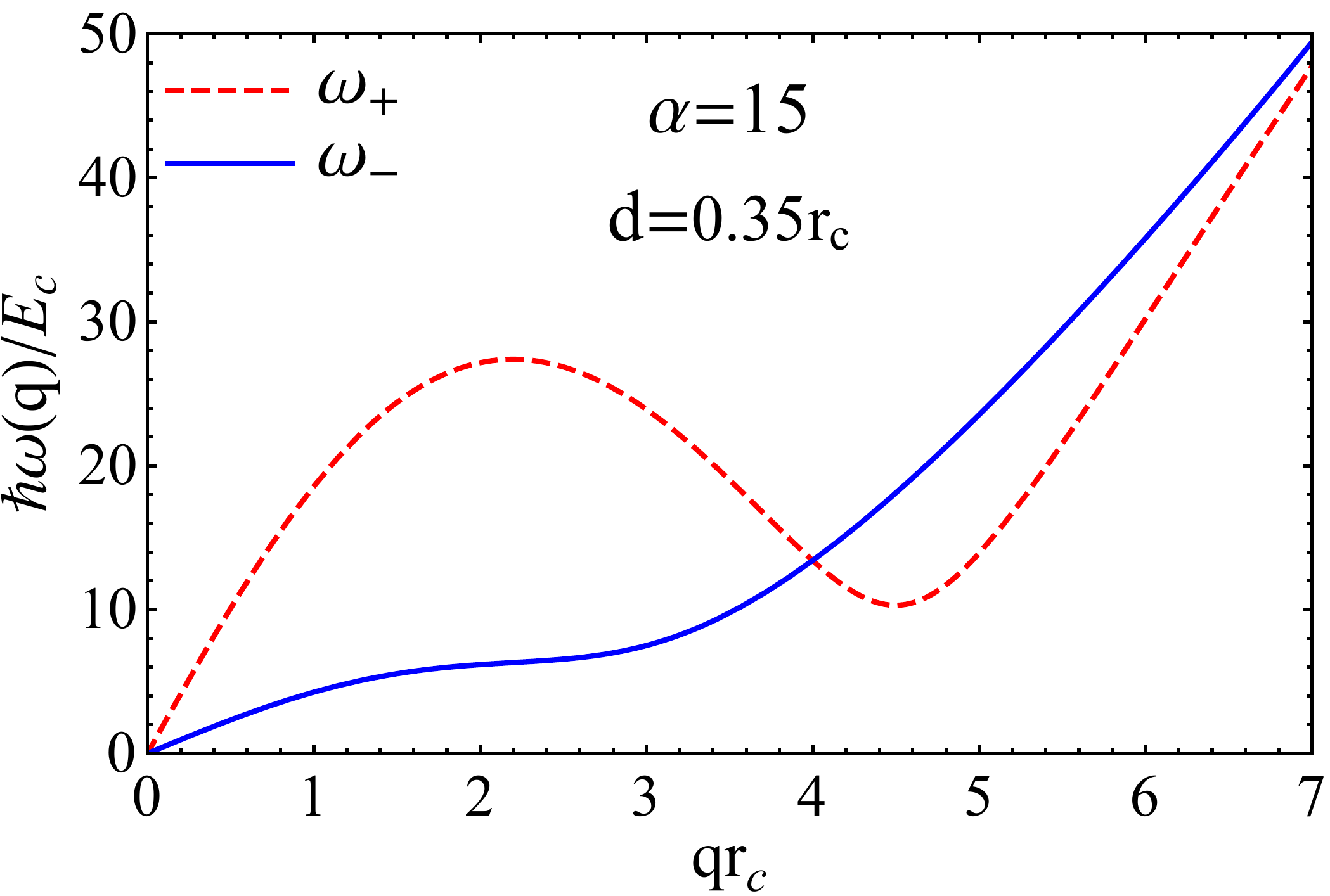}&
		\includegraphics[width=0.5\linewidth]{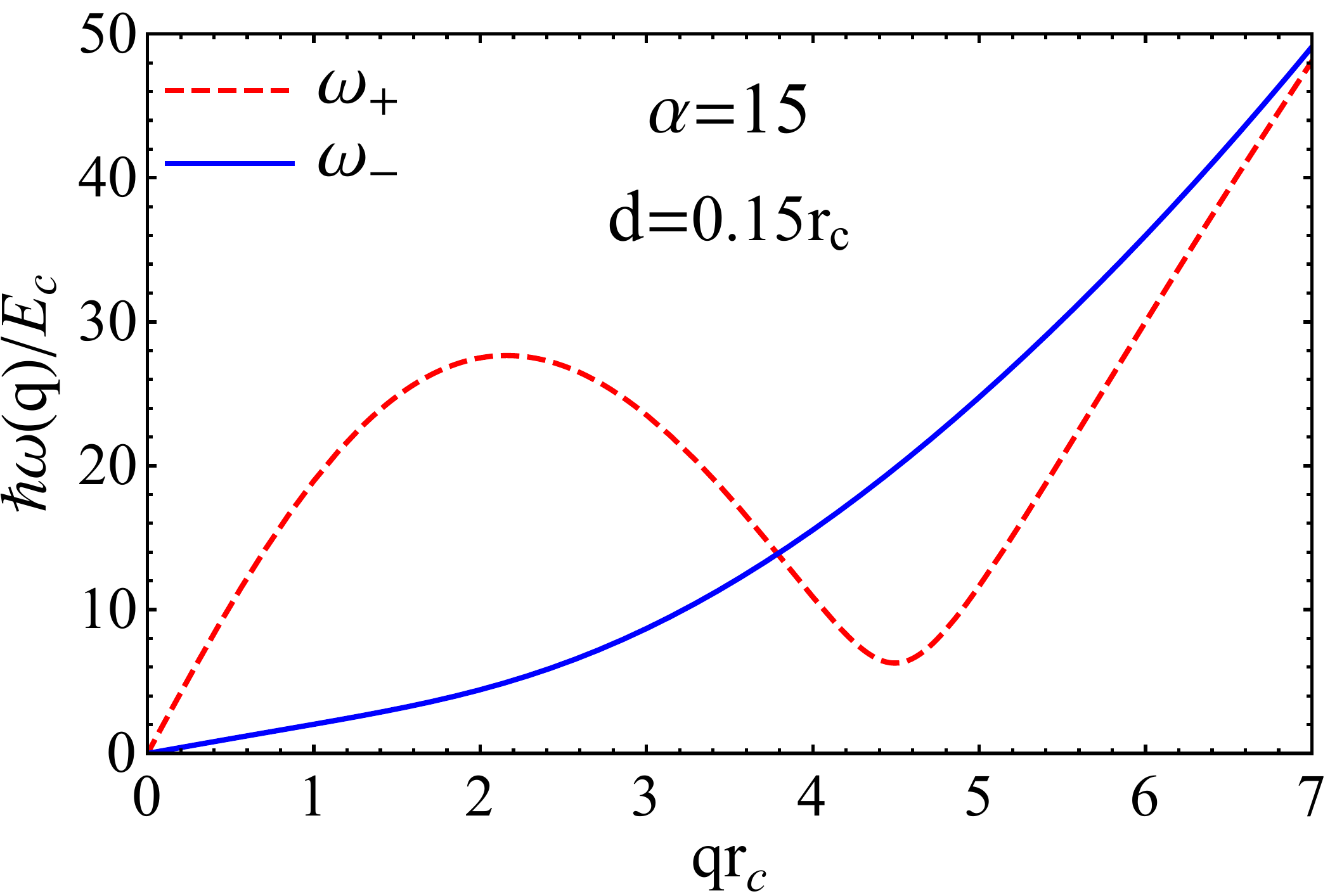}
	\end{tabular}
	\caption{Dispersions of density ${\omega _ + }$ and pseudospin ${\omega _ - }$ modes of a Rydberg-dressed Bose bilayer [in units of $\hbar^2/(2mr_c^2)$] for different values of the interlayer distances. Here we use $\alpha = 15$, which is close to the density-wave instability region at small layer spacings.
	\label{fig:w_pm_rydberg}}
\end{figure}

\begin{figure}
	\begin{tabular}{cc}
		\includegraphics[width=0.5\linewidth]{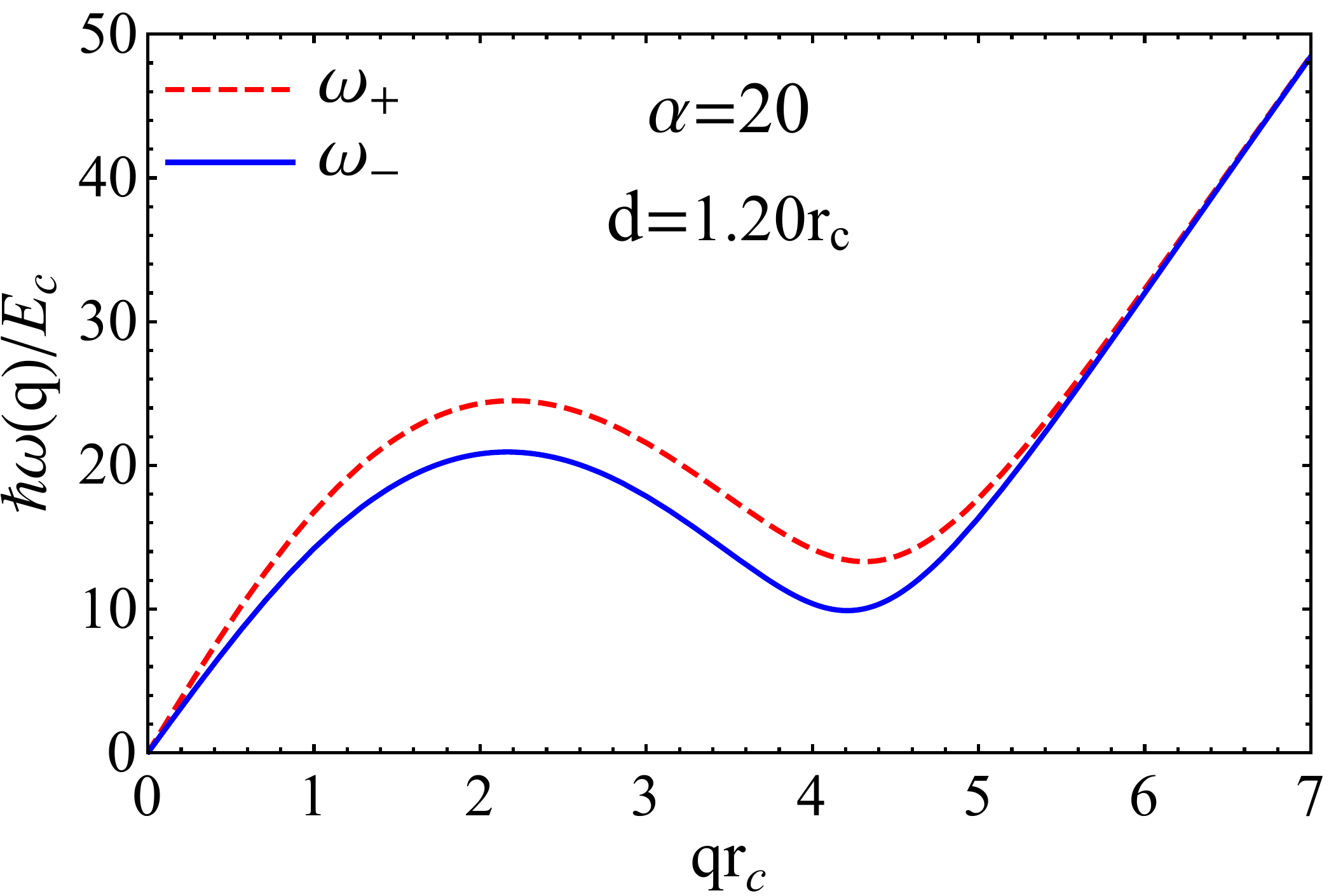}&
		\includegraphics[width=0.5\linewidth]{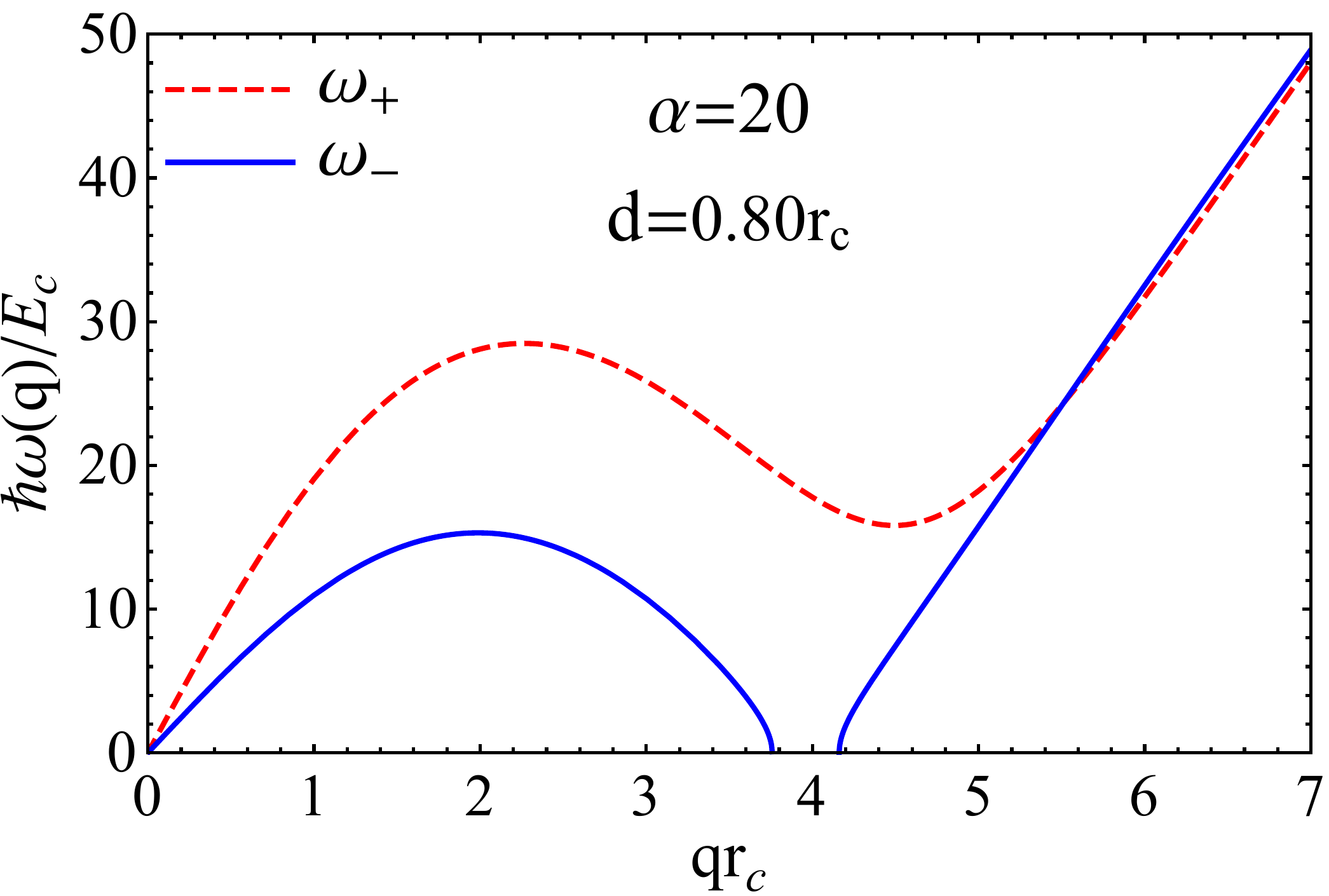}\\
		\includegraphics[width=0.5\linewidth]{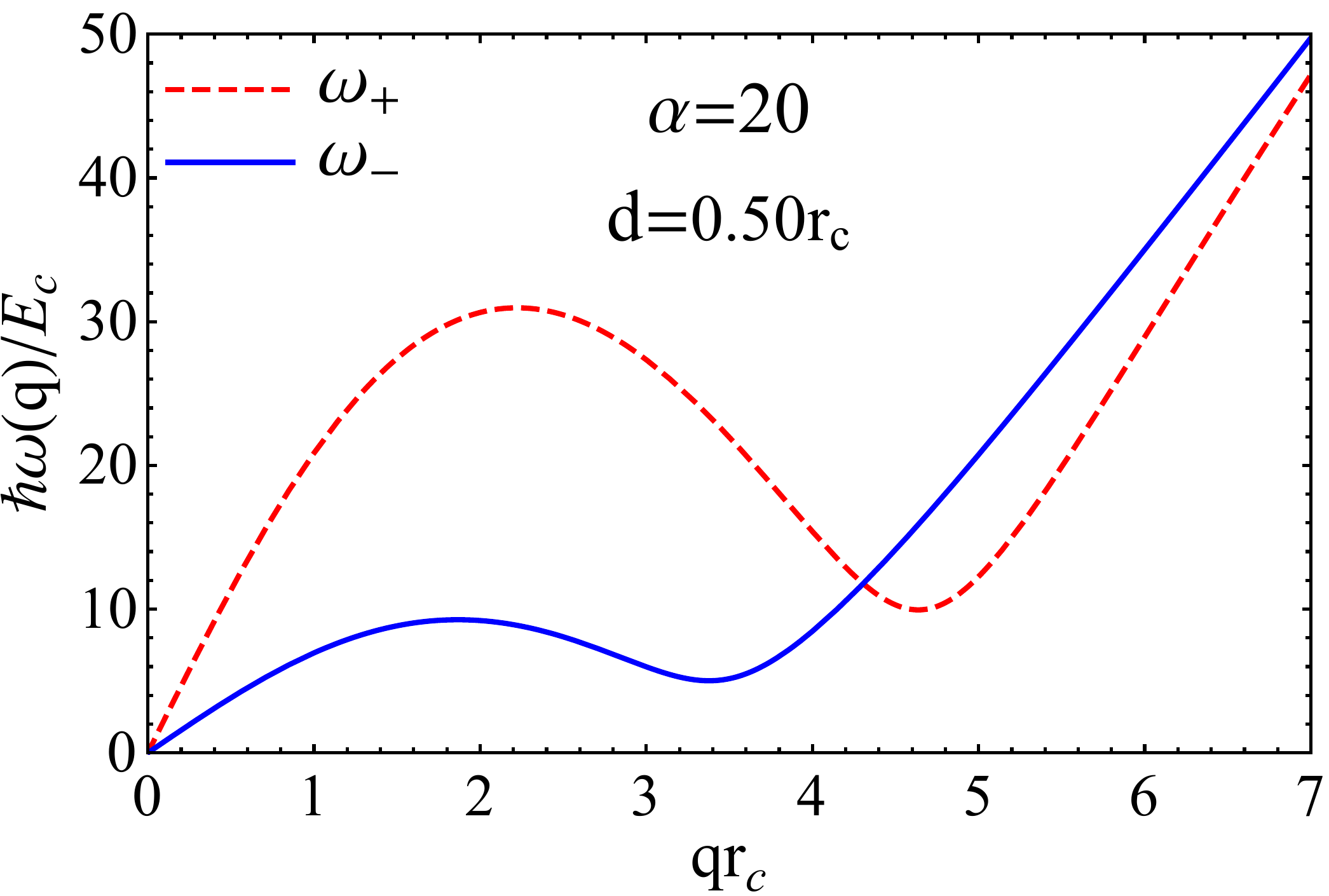}&
		\includegraphics[width=0.5\linewidth]{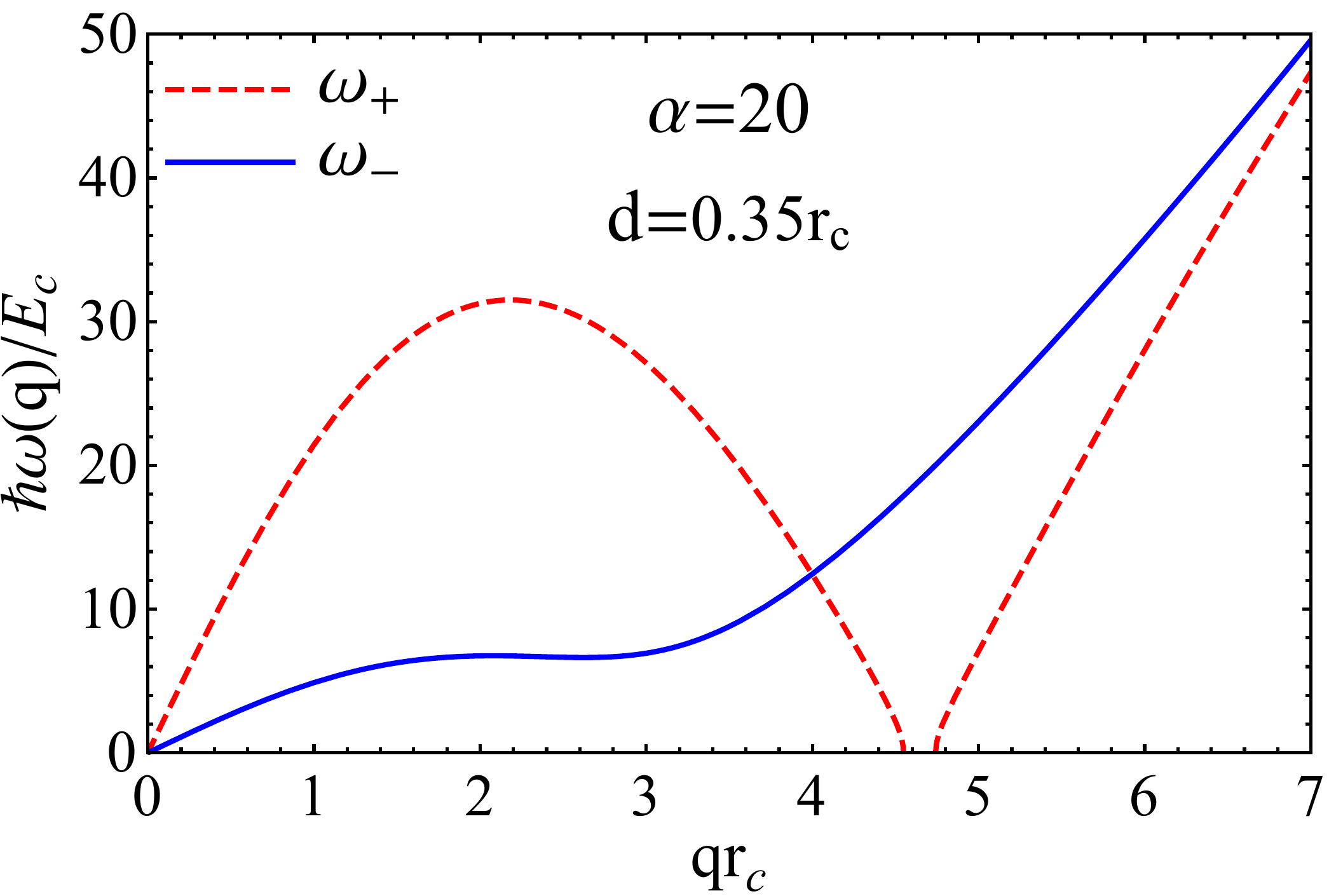}
	\end{tabular}
	\caption{Similar to Fig.\,\ref{fig:w_pm_rydberg} but for $\alpha = 20$. The vanishing of the roton energy is evident in the right panels.
	\label{fig:w_pm_rydberg2}}
\end{figure}

In Figs.~\ref{fig:w_pm_rydberg} and~\ref{fig:w_pm_rydberg2} we show the dispersion of both symmetric and asymmetric collective modes in the bilayer system of Rydberg-dressed bosons. In each figure, we keep the coupling constant $\alpha$ fixed but vary the layer spacing to illustrate the effect of interlayer coupling on the mode dispersion.  
Note that at large layer spacings (i.e., $d \gg r_c$), the two layers are almost decoupled and the two collective mode branches become degenerate.
For the values of $\alpha$ we have chosen here, the roton minima are already visible in the density mode 
$(\omega_+)$ of two decoupled layers.
As we decrease the distance between the two layers, the asymmetric mode $(\omega_-)$ becomes softer. 
At smaller distances, the asymmetric component of the interaction $V_-(q)$ gets smaller and the dispersion of the pseudo-spin mode becomes free-particle-like (see, e.g. the bottom right panel of 
Fig.\,\ref{fig:w_pm_rydberg}). The roton in the total density mode becomes very soft in this regime, almost similar to the density mode of a single layer with the coupling constant of $2\alpha$.
The energy of both roton minima changes non-monotonically with the layer spacing. Vanishing of the roton minima for $\alpha=20$ in Fig.\,\ref{fig:w_pm_rydberg2} for  ${\tilde d}=0.8$ (top right panel) and ${\tilde d}=0.35$ (bottom right panel) indicate density-wave instability in the asymmetric and symmetric channels, respectively. 

Another interesting observation is that two modes become degenerate at finite wave vectors where $V_{\rm d}(q)=0$. This degeneracy breaks when the symmetry between two layers is broken through \emph{e.g.} density imbalance or a finite counterflow, as we will discuss in the following subsections.


\subsection{Density-wave instability}\label{DW}
As we have seen in Fig.\,\ref{fig:w_pm_rydberg2}, at strong coupling strengths, the energy of roton approaches zero, and eventually, the energy of quasiparticle excitations can become imaginary. This signals the instability of the homogeneous superfluid phase towards density waves for some ranges of the layer spacing. In this subsection, we further explore this phenomenon.

\begin{figure}
	\begin{tabular}{c}
		\includegraphics[width=\linewidth]{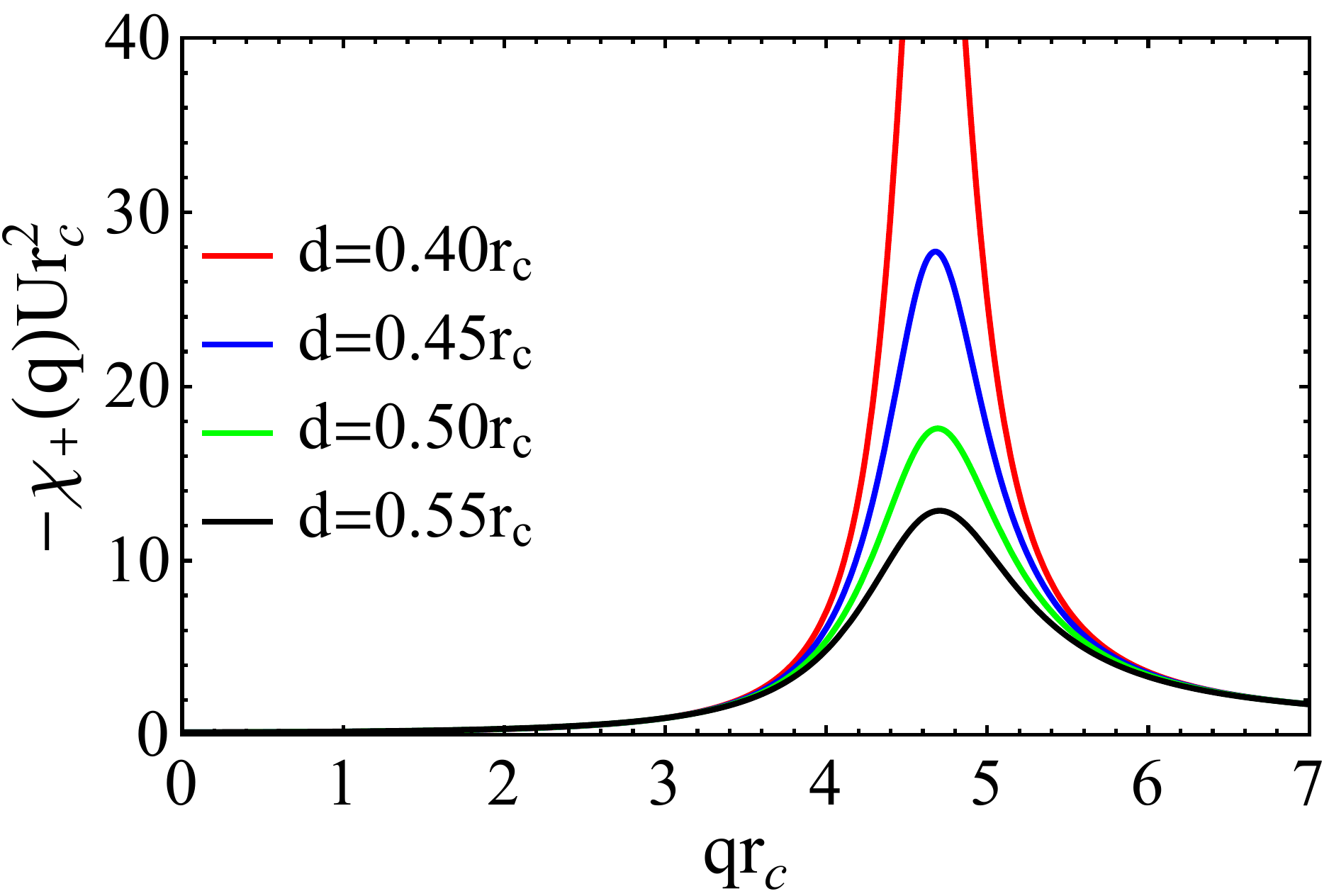}\\
		\includegraphics[width=\linewidth]{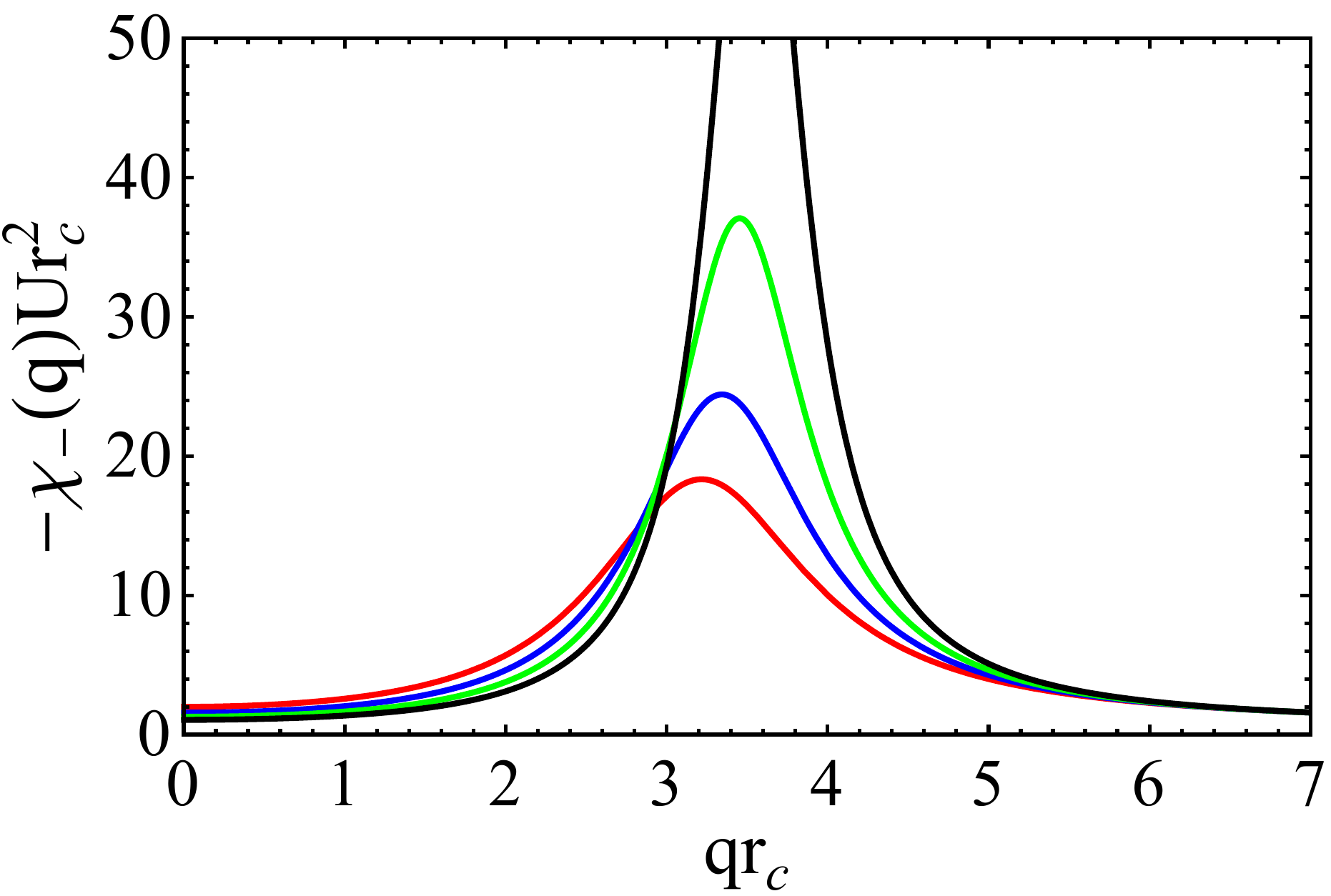}
	\end{tabular}
	\caption{The symmetric (top) and asymmetric (bottom) components of the density-density response function [in the units of $1/(Ur_c^2)$] within the random phase approximation versus $q r_c$ for $\alpha =20$.
		\label{fig:response_func_rydberg}}
\end{figure}

The linear density-density response function matrix of a bilayer system could be written as
 \be
	{\hat \chi} (q,\omega ) = {\left[ {I - \hat \Pi (q,\omega ){{\hat W}^{\rm eff}}(q,\omega )} \right]^{ - 1}}\hat \Pi (q,\omega ),
 \ee
where $I$ and ${\hat W^{\rm eff}}(q,\omega )$ are the $2\times 2$ identity and effective potential matrices, respectively. Within the random phase approximation, the dynamical effective potential is replaced by the bare potentials, whose matrix elements are
 \be
 {V_{ij}}(q) = {\delta _{ij}}{V_{\rm s}}(q) + (1 - {\delta _{ij}}){V_{\rm d}}(q),
 \ee 
and $\hat \Pi (q,\omega )$ is the zero temperature non-interacting density-density response matrix of a Bose gas, whose elements for a symmetric bilayer are
$ {\Pi_{ij}}(q,\omega) = {\delta _{ij}}\Pi(q,\omega)$, 
 with
 \be
	\Pi (q,\omega ) = \frac{{2n{\varepsilon _q}}}{{{{(\hbar \omega  + i{0^ + })}^2} - \varepsilon _q^2}}.
 \ee
For a symmetric bilayer structure, the linear density-density response matrix could be diagonalized into the symmetric and asymmetric components~\cite{abedinpour_LTP2020}
  \be
    \chi _ \pm (q,\omega ) = \frac{{\Pi (q,\omega )}}{{1 - \Pi (q,\omega ){V_ \pm }(q)}}.
  \ee
Furthermore, the screened potential in the matrix form, is also given as
 \be
	{\hat V^{\rm sc}}(q,\omega ) = {\left[ {I - \hat \Pi (q,\omega ){{\hat W}^{\rm eff}}(q,\omega )} \right]^{ - 1}}{\hat W^{\rm eff}}(q,\omega ).
 \ee
In the static limit, the intralayer and interlayer screened potentials read~\cite{moudgil}
\be
\begin{split}
	V_{\rm s}^{\rm sc}(q) =& \frac{{{V_{\rm s}}(q) + [V_{\rm d}^2(q) - V_{\rm s}^2(q)]\Pi(q) }}{{{{[1 - \Pi(q) {V_{\rm s}}(q)]}^2} - {\Pi ^2}(q)V_{\rm d}^2(q)}},\\
	V_{\rm d}^{\rm sc}(q) =& \frac{{{V_{\rm d}}(q)}}{{{{[1 - \Pi(q) {V_{\rm s}}(q)]}^2} - {\Pi ^2}(q)V_{\rm d}^2(q)}},
\end{split}
\ee
where $\Pi(q)\equiv\Pi(q,\omega=0)=-2n/\varepsilon_q$ is the static limit of the non-interacting density-density response function. 
In Fig.\,\ref{fig:response_func_rydberg} we illustrate the wavevector dependence of the symmetric and asymmetric components of the response function for several values of the layer spacings. We observe that the symmetric component of the response function diverges at larger layer spacing while this behavior appears in the asymmetric component at smaller layer spacings. This implies that the density-wave instability could occur in both of these channels.  

\begin{figure}
	\begin{tabular}{c}
		\includegraphics[width=\linewidth]{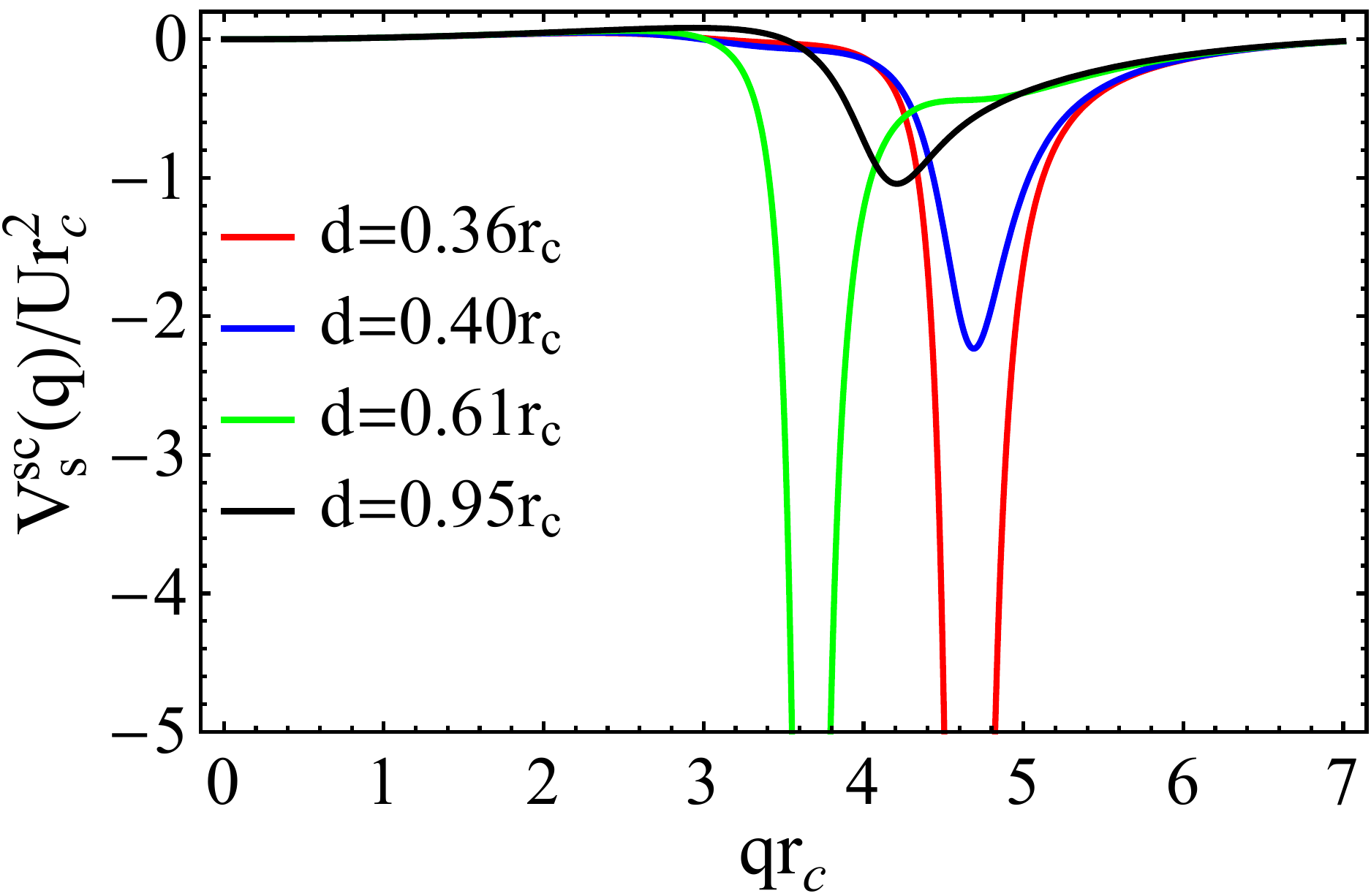}\\
		\includegraphics[width=\linewidth]{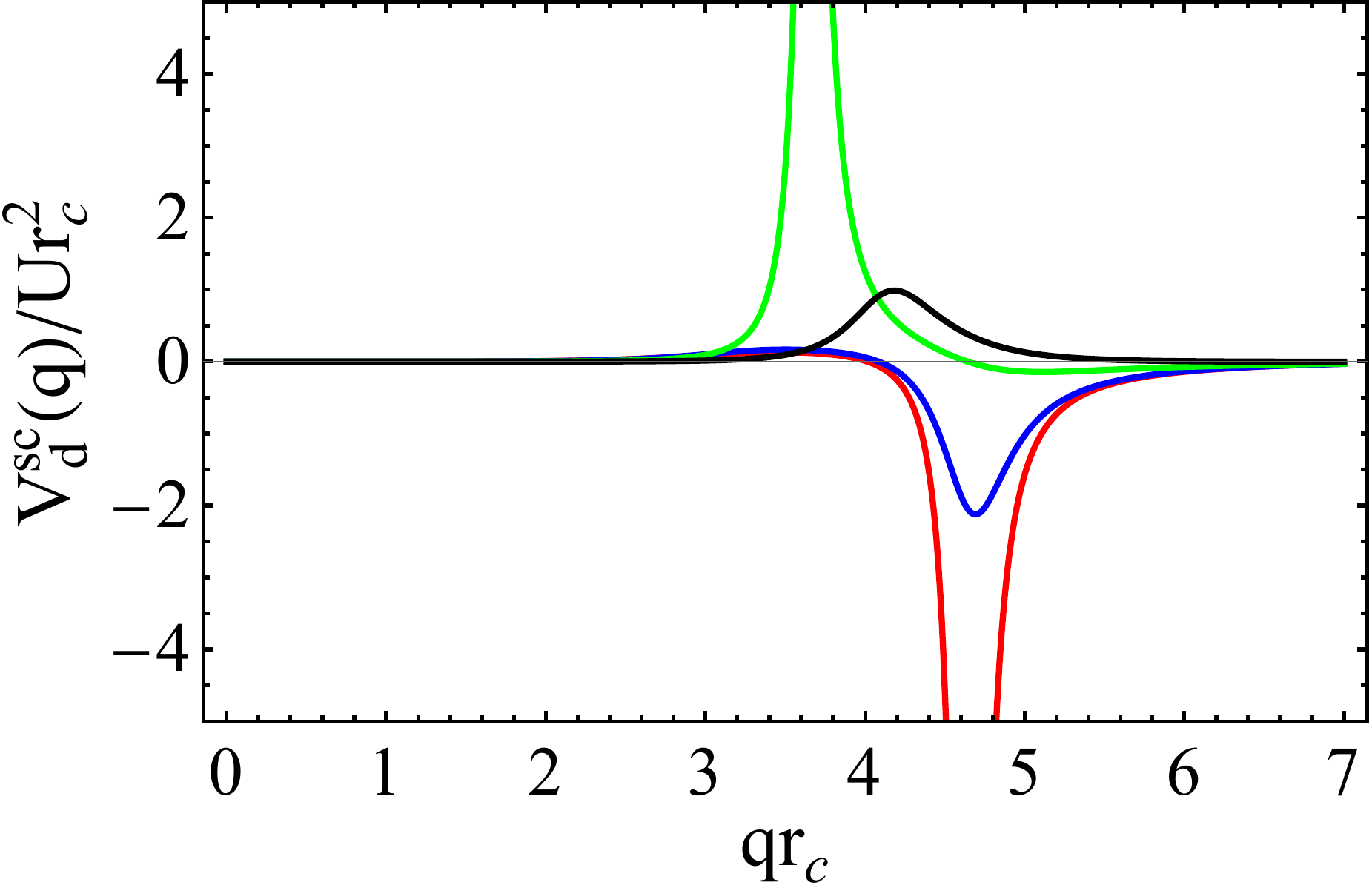}
	\end{tabular}
	\caption{The intralayer (top) and interlayer (bottom) screened potentials versus $qr_c$ for different values of the layer spacings and for $\alpha =20$.
		\label{fig:Vq_Sc_rydberg}}
\end{figure}

We also investigate the effect of the interlayer separation on the screened intralayer and interlayer potentials in 
Fig.\,\ref{fig:Vq_Sc_rydberg}. As the layer spacing approaches specific values, sharp peaks emerge in both components of the screened potential for $\alpha =20$. This indicates that the homogeneous bilayer system becomes unstable towards density modulation phases. 
It is also noted that two components of the screened potential remain finite at weaker coupling constant (i.e., for $\alpha=15$), as one would expect from the excitation spectrums of Fig.\,\ref{fig:w_pm_rydberg}.


\begin{figure}
	\includegraphics[width=\linewidth]{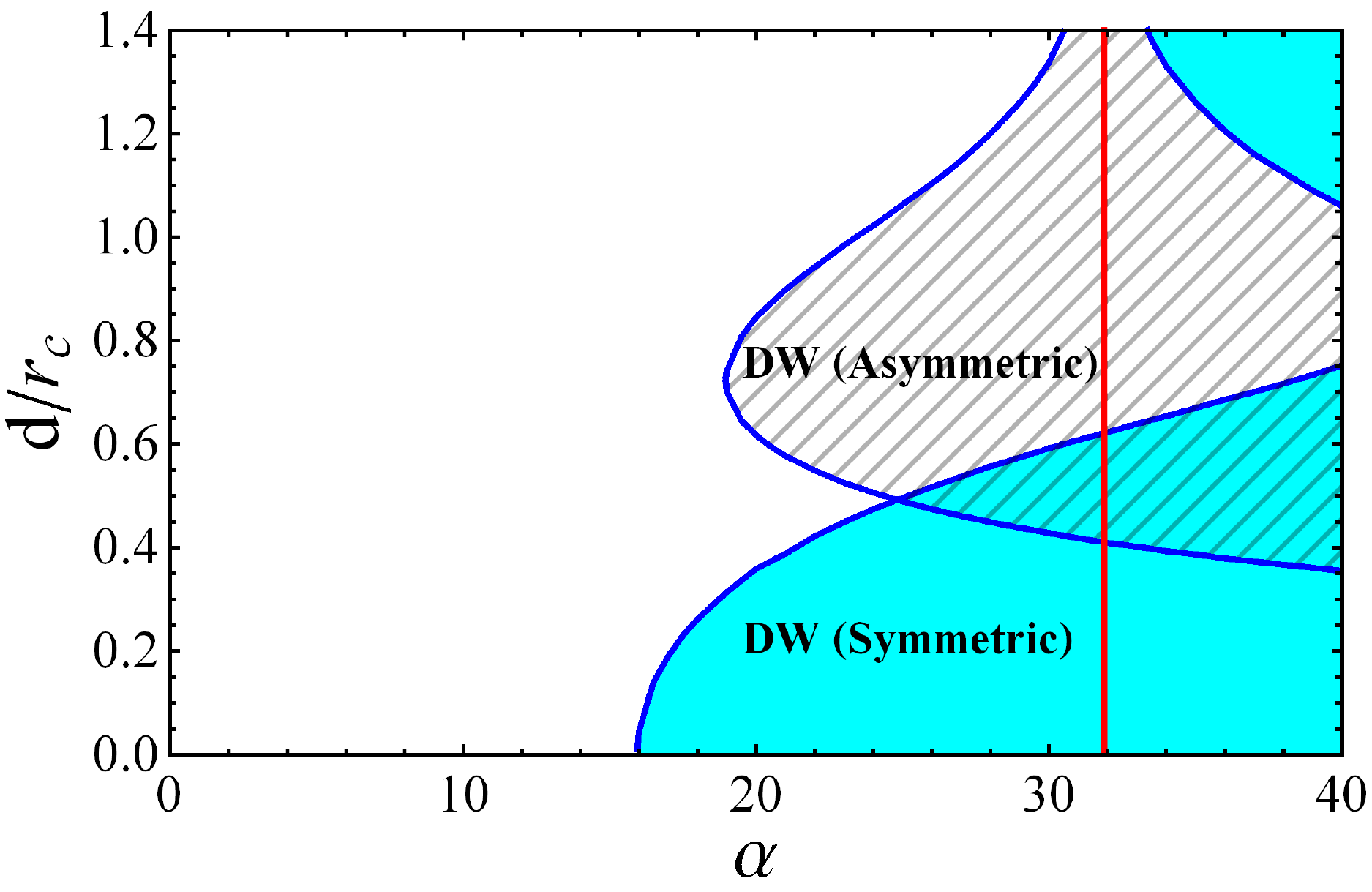}
	\caption{The phase diagram of a bilayer of Bose gas with Rydberg-dressed interaction in the $d$-$\alpha$ plane. The homogeneous superfluid (SF), and density-wave (DW) instability in the symmetric and asymmetric channels are marked. The vertical red line at $\alpha_c\approx 31.88$ is the critical coupling strength for the density-wave instability of an isolated two-dimensional layer of Rydberg-dressed bosons.	
	\label{fig:phase_DW}}
\end{figure}

Singularities of the density-density response function and the screened potentials are signatures of the 
density wave instability.
We have investigated a Bose bilayer with Rydberg-dressed interactions for such instabilities in both symmetric (i.e., density) and asymmetric (i.e., pseudo-spin) channels. The phase diagram as shown in Fig.\,\ref{fig:phase_DW} contains homogeneous superfluid (SF), density-wave (DW) instabilities in either symmetric or asymmetric channels, and the region where instability appears in both of these channels.


\subsection{Counterflow}\label{sec:2}
\begin{figure}
	\begin{tabular}{c}
		\includegraphics[width=\linewidth]{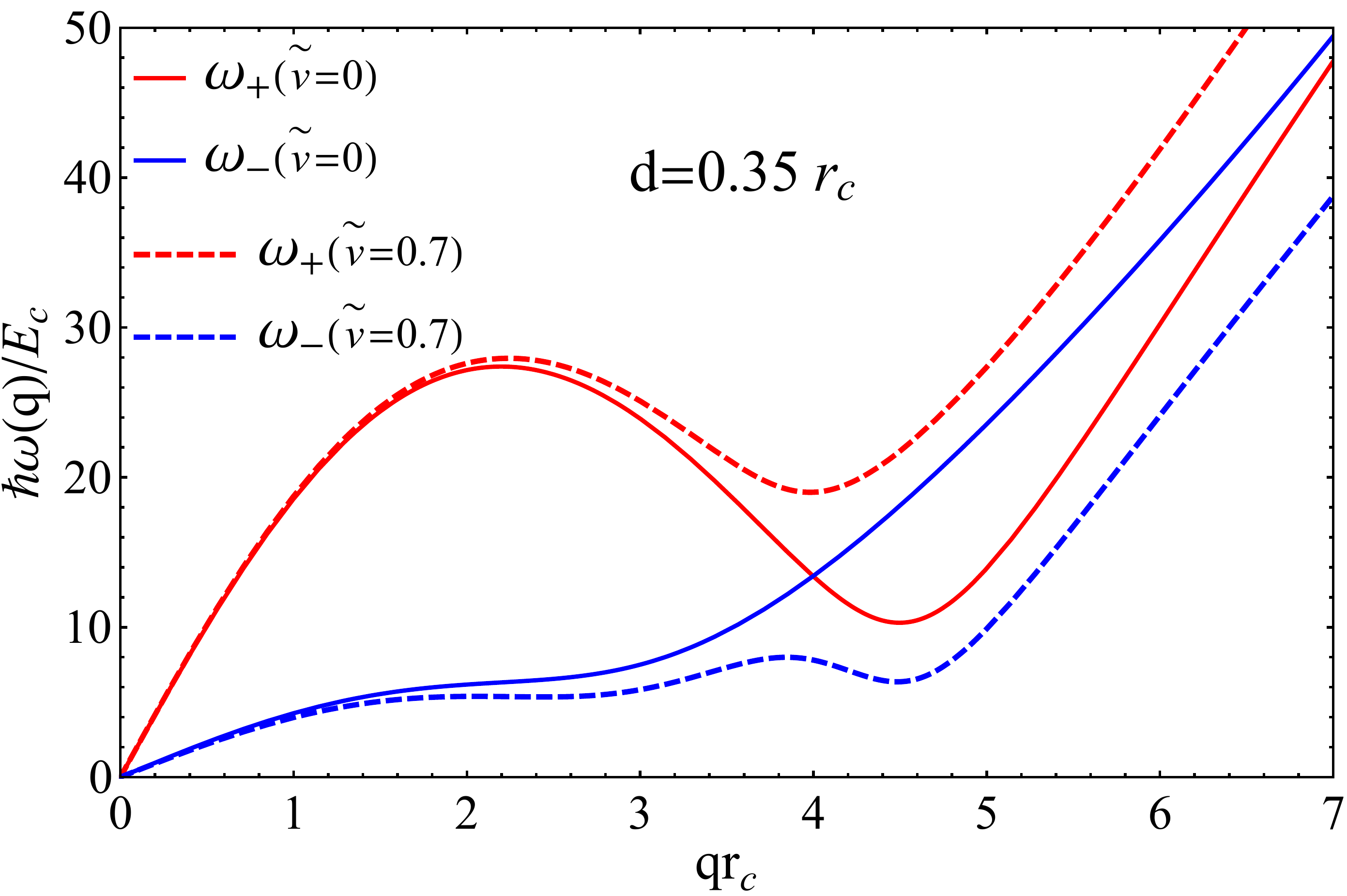}\\
		\includegraphics[width=\linewidth]{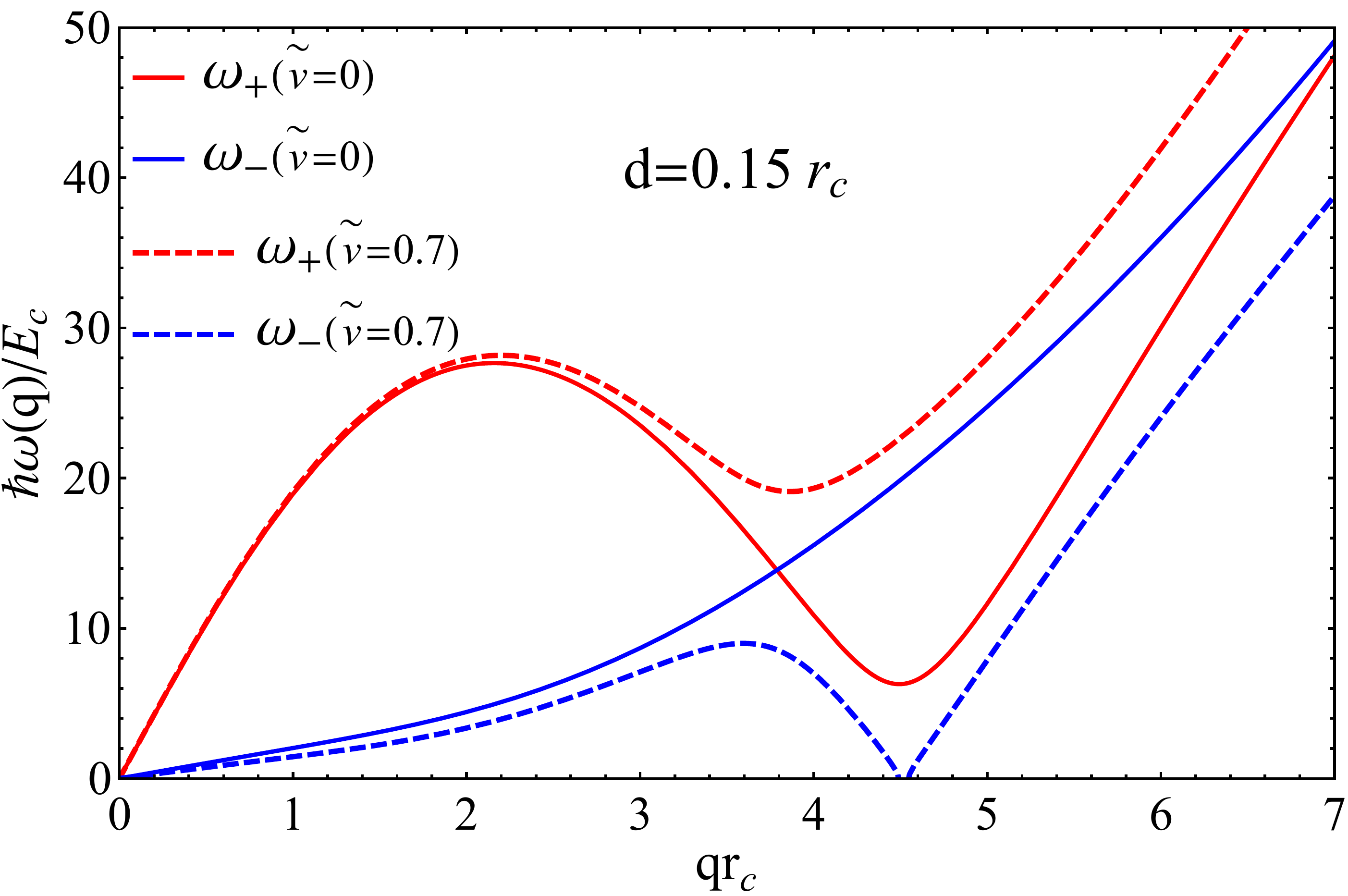}
	\end{tabular}
\caption{Dispersions of the density and pseudo-spin collective modes of a Rydberg-dressed Bose bilayer at finite counterflow ${\tilde v}=m r_c v/\hbar=0.7$, and for ${\bf q}||{\bf v}$. The dimensionless coupling constant is fixed at $\alpha =15$ in both panels.  		\label{fig:p_counterflow_rydberg}}
\end{figure}

If we include finite background velocities in the first and second layers, denoted by ${{\bf{v}}_1}$  and ${{\bf{v}}_2}$, respectively, the collective modes could be obtained by finding the poles of the total density-density response function in which $\omega$  is replaced by $\omega  - {{\bf{v}}_i} \cdot {\bf{q}}$  in layer $i=1, 2$. For unequal values of velocities, the background flow could be decomposed into the center-of-mass ${\bf{V}} = ({{\bf{v}}_1} + {{\bf{v}}_2})/2$ and counterflow  ${\bf{v}} = ({{\bf{v}}_1} - {{\bf{v}}_2})/2$ components. We neglect the center-of-mass flow since its effect could be found in terms of a Galilean boost. Then, we concentrate on the counterflow term and solve the following determinant equation~\cite{abedinpour_LTP2020}.
\be\label{eq:w_pm_eq}
\left| {\begin{array}{*{20}{c}}
	{{\Pi ^{ - 1}}(q,\omega  - {\bf{v}} \cdot {\bf{q}}) - {V_{\rm s}}(q)}&{ - {V_{\rm d}}(q)}\\
	{ - {V_{\rm d}}(q)}&{{\Pi ^{ - 1}}(q,\omega  + {\bf{v}} \cdot {\bf{q}}) - V_{\rm s}(q)}
	\end{array}} \right| = 0.
\ee
In the presence of finite counterflow, the dispersion of collective modes are straightforwardly obtained as 
\be\label{eq:w_pm_v_full}
\begin{split}
 	{\hbar ^2}\omega _ \pm ^2  =& \varepsilon _q^2\left[ {1 + \frac{{2n}}{{{\varepsilon _q}}}V_{\rm s}(q)} \right] + {(\hbar {\bf{v}} \cdot {\bf{q}})^2}  \\
 	&\pm 2{\varepsilon _q}\sqrt {{n^2}V_{\rm d}^2(q) + {{(\hbar {\bf{v}} \cdot {\bf{q}})}^2}\left[ {1 + \frac{{2n}}{{{\varepsilon _q}}}{V_{\rm s}}(q)} \right] },
	\end{split}
\ee
which is evidently anisotropic. For small counterflow velocities, we can write 
 \be\label{eq:w_pm_v}
 \hbar {\omega _ \pm }({\bf{q}},{\bf{v}}) \approx \hbar {\omega _ \pm }({\bf{q}}) + \frac{1}{2}{M_ \pm }(q,\phi ){v^2},
 \ee
 where
  \be\label{eq:w_pm_M}
 {M_ \pm }(q,\phi ) = \frac{\hbar q^2 {\cos}^2(\phi)}{{{\omega _ \pm }(q,0)}}\left\{ {1 \pm \frac{{{\varepsilon _q}}}{{n{V_{\rm d}}(q)}}\left[ {1 + \frac{{2n}}{{{\varepsilon _q}}}{V_{\rm s}}(q)} \right]} \right\}.
\ee
Here, ${\omega _ \pm }(q,0)$ the dispersions of collective modes in the absence of counterflow are given through Eq.~\eqref{eq:w_pm} and $\phi$  is the angle between the wave vector ${\bf q}$ and the counterflow direction. 
Figure~\ref{fig:p_counterflow_rydberg} shows the dispersion of collective modes at a finite counterflow velocity obtained from Eq.~\eqref{eq:w_pm_v_full}. 
Counterflow raises the energy of the high-energy branch and makes the low energy mode softer. In other words, a finite counterflow can drive a homogeneous system towards the density-wave instability (see, the bottom right panel in Fig.\,\ref{fig:p_counterflow_rydberg}).

\subsection{Density imbalanced bilayer}
In this section, we consider a density imbalanced bilayer of Rydberg-dressed bosons where the particle density is different in two layers. 
The collective modes of this asymmetric bilayer could be readily obtained as
\be\label{eq:w_pm_im}
\hbar {\omega _ \pm }(q)=  \sqrt{\varepsilon _q^2+ 2n{\varepsilon _q}\left[ V_{\rm s}(q) \pm \sqrt {V_{\rm d}^2(q) +p^2V_+(q)V_-(q) } \right]},
\ee
where $n=(n_1+n_2)/2$ is now the average density, and $p=(n_1-n_2)/(n_1+n_2)$ is the density polarization. 
Eq.\,\eqref{eq:w_pm_im} reduces to Eq.\,\eqref{eq:w_pm} for $p=0$.

\begin{figure}
	\begin{tabular}{c}
		\includegraphics[width=\linewidth]{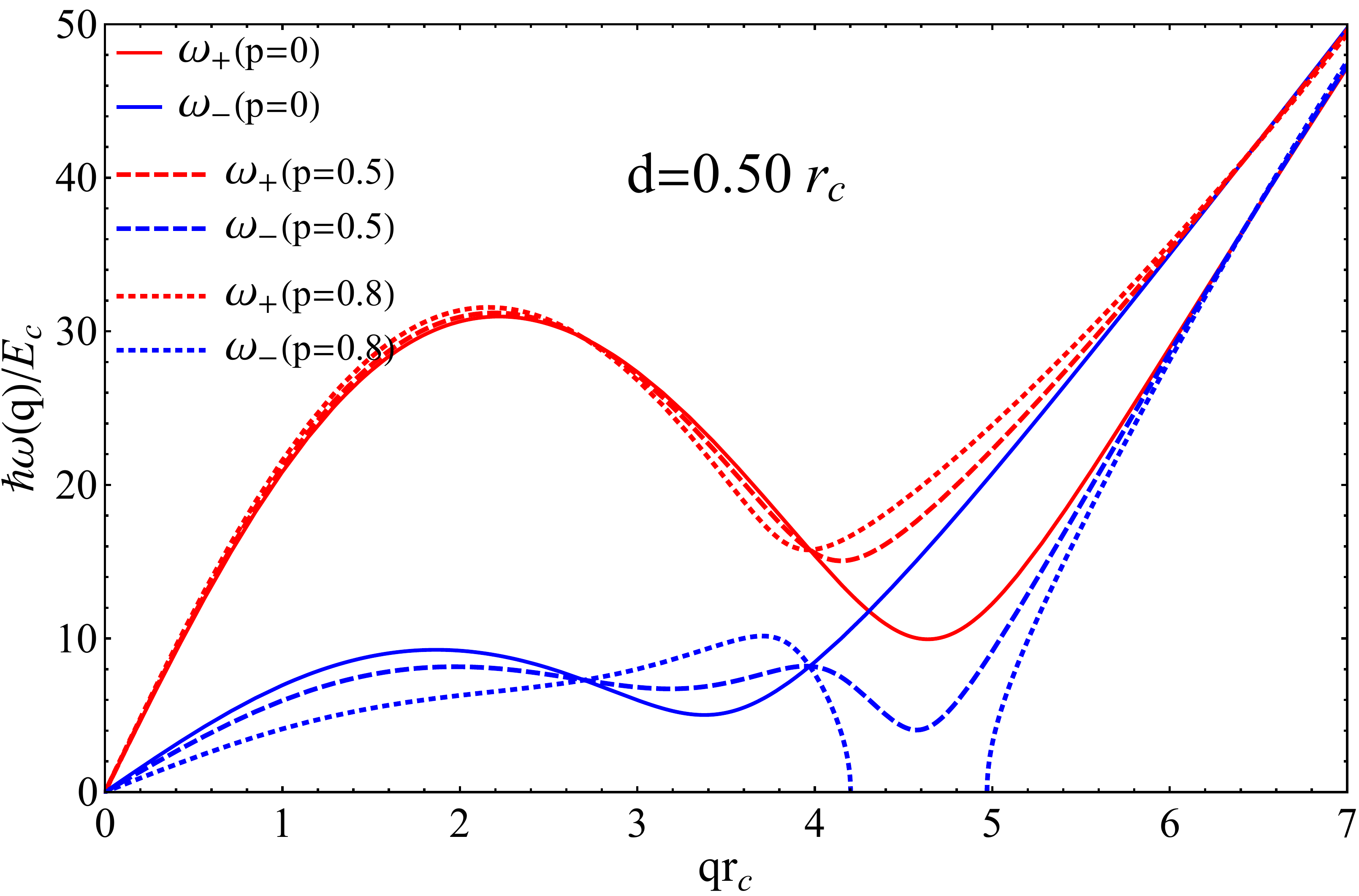}\\
		\includegraphics[width=\linewidth]{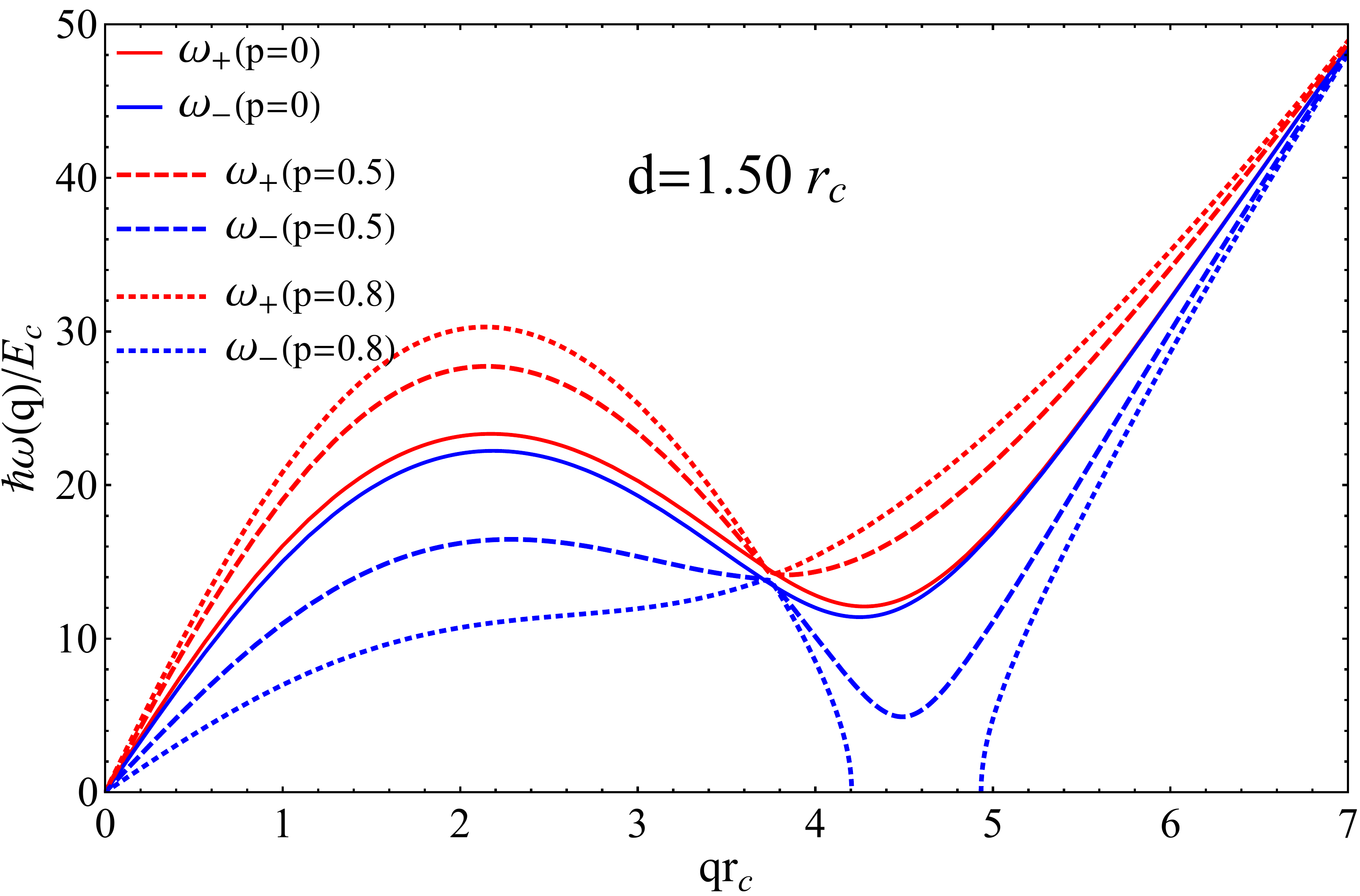}
	\end{tabular}
	\caption{Dispersions of two modes of a density imbalanced Rydberg-dressed Bose bilayer at $\alpha =20$ with different polarizations and different layer separations.}
	\label{fig:w_pm_density}
\end{figure}

In Fig.\,\ref{fig:w_pm_density} the dispersion of collective modes of a density imbalanced bilayer is presented. Similar to the counterflow, density imbalance makes the system more susceptible for the density-wave instability.
\subsection{The density-of-states of collective modes\label{DOS}}
\begin{figure}
	\begin{tabular}{cc}
		\includegraphics[width=0.5\linewidth]{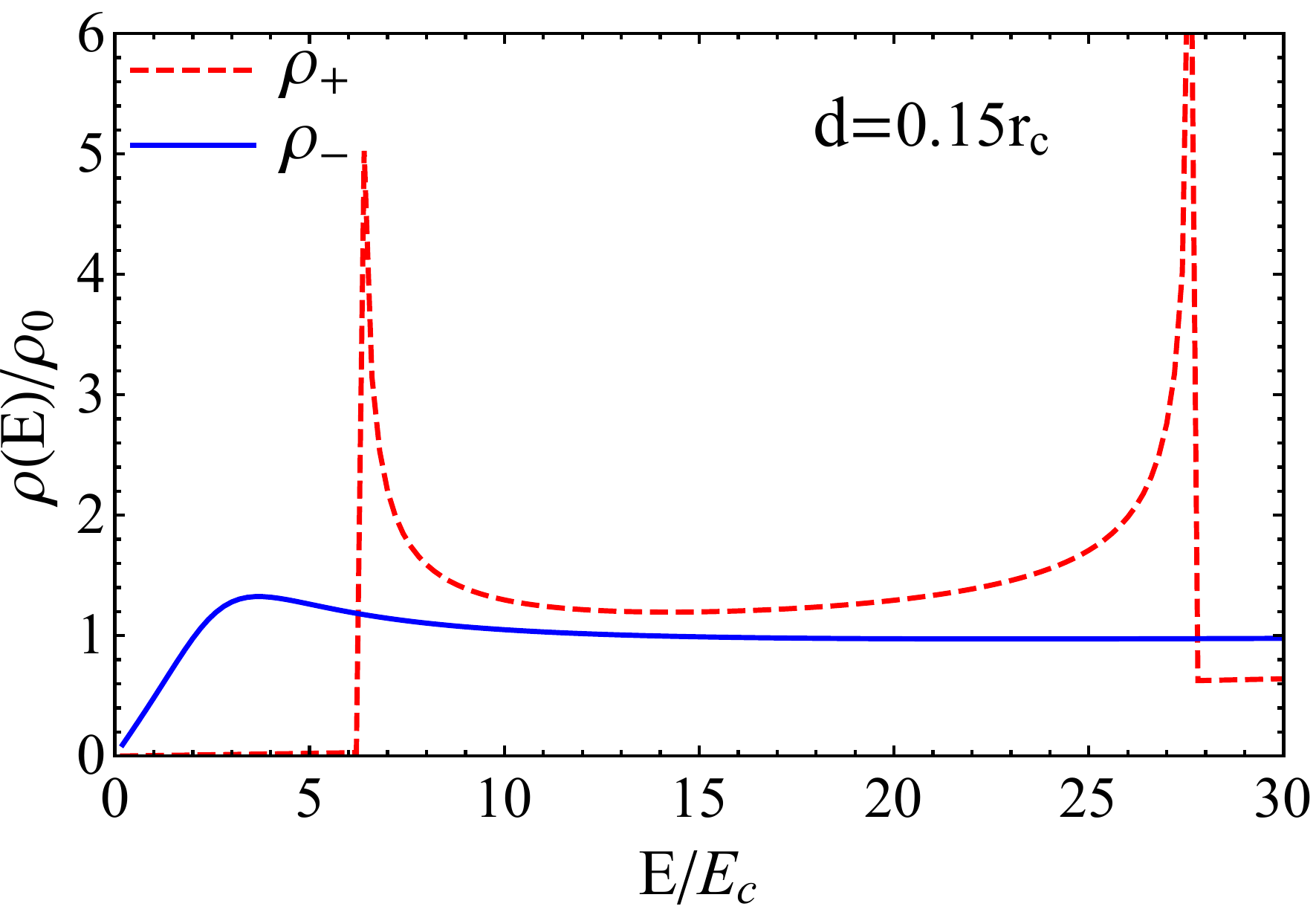}&
		\includegraphics[width=0.5\linewidth]{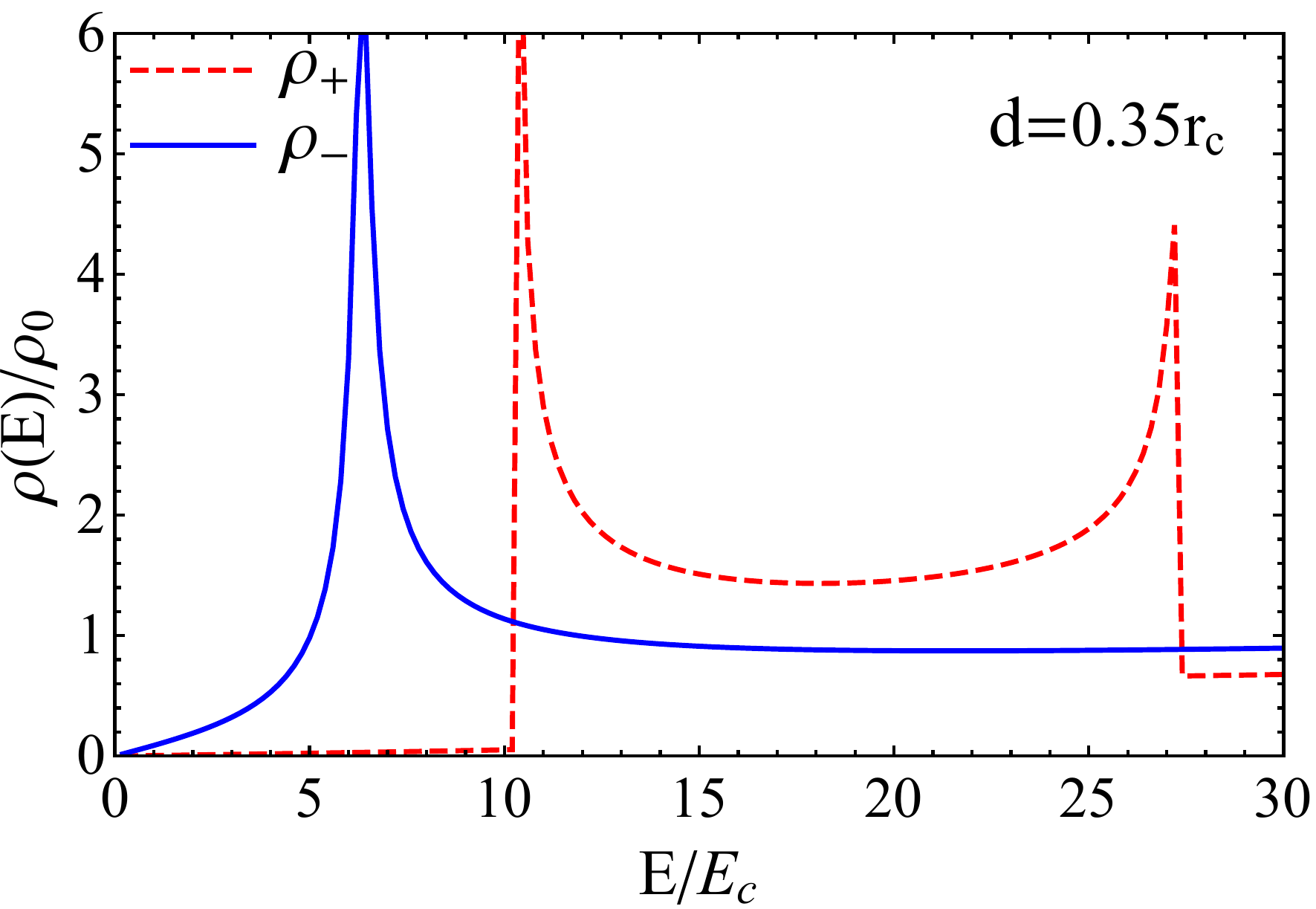}\\
		\includegraphics[width=0.5\linewidth]{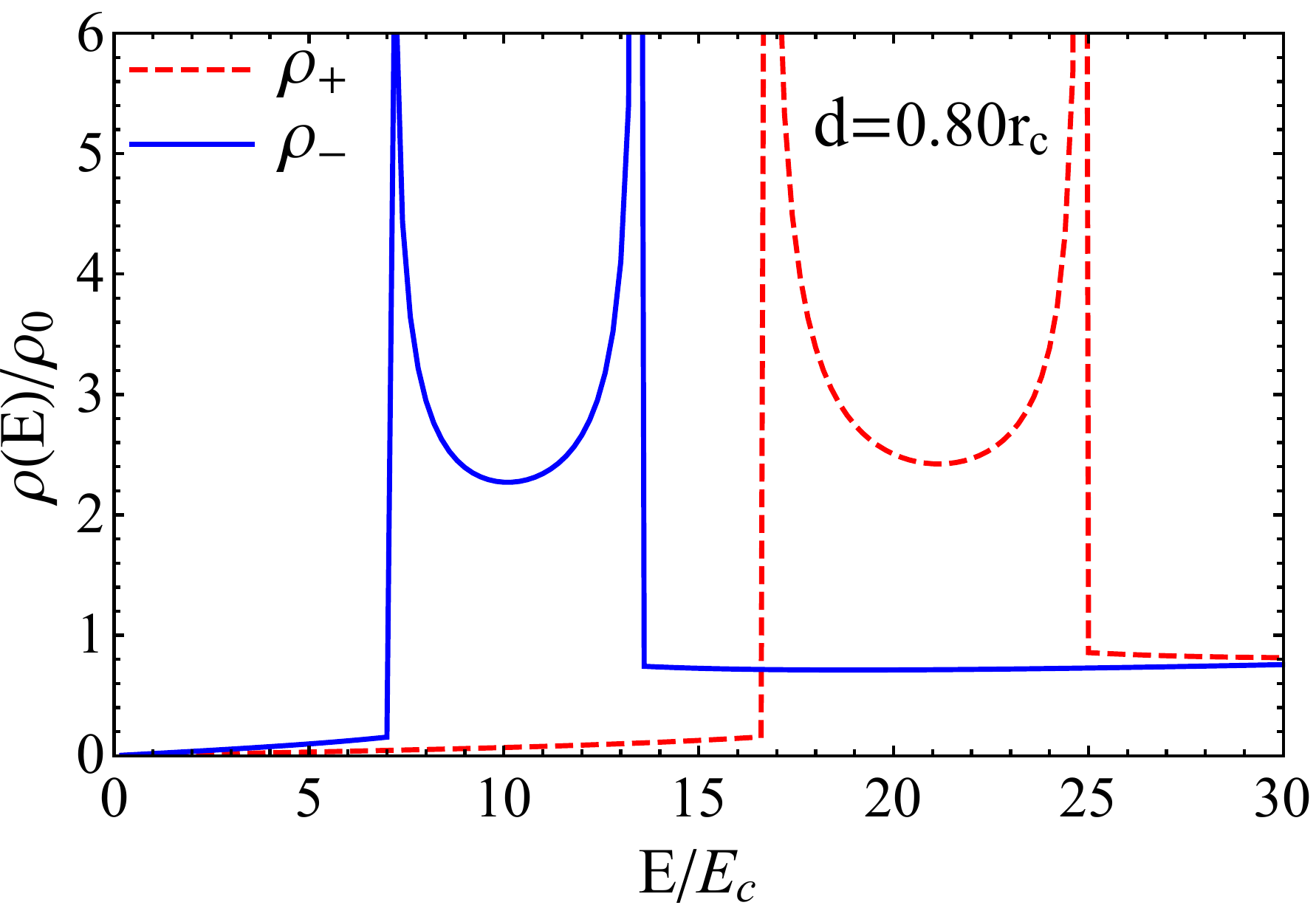}&
		\includegraphics[width=0.5\linewidth]{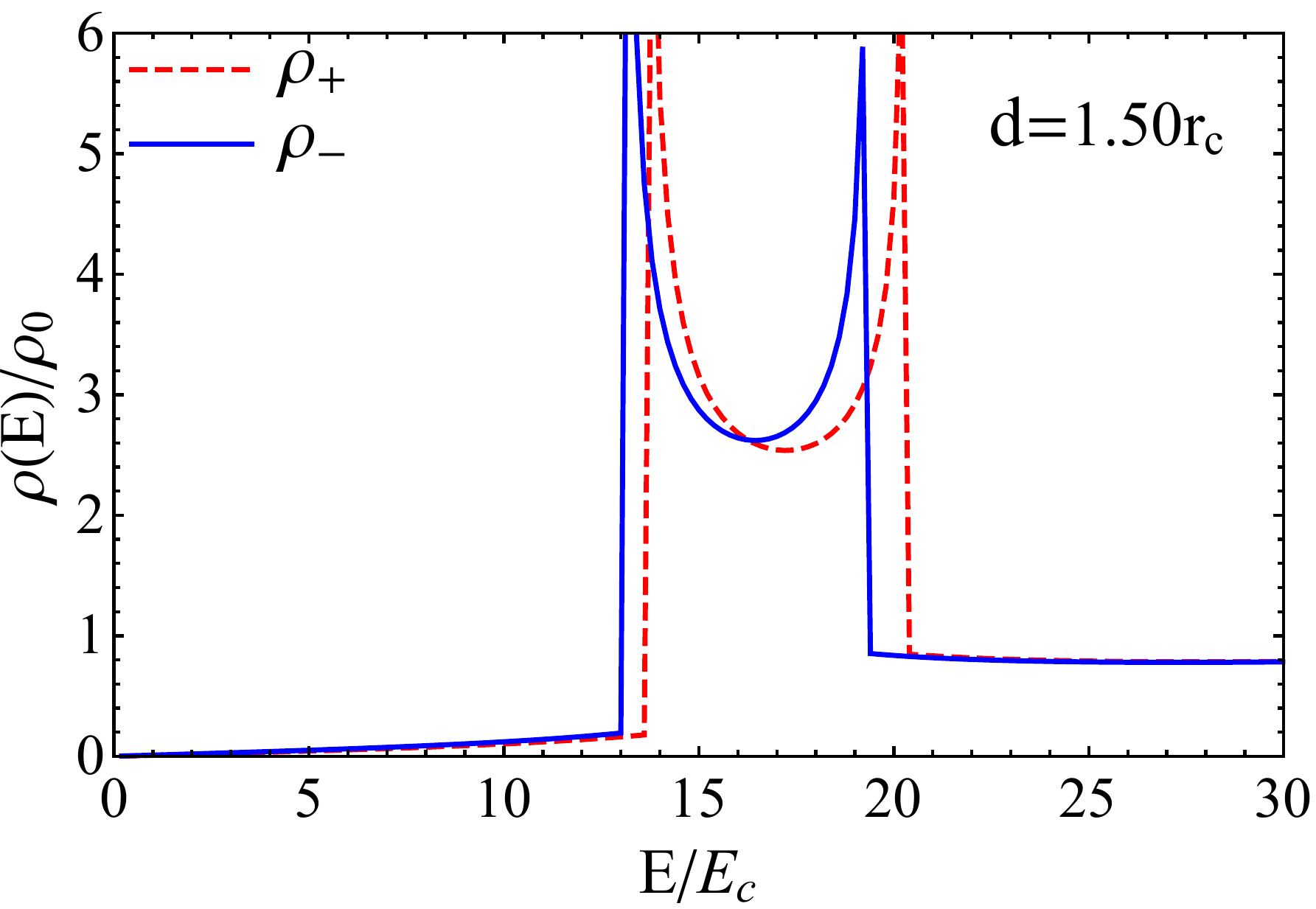}
	\end{tabular}
	\caption{Density-of-states per unit area [in the units of $\rho_0=m/(2\pi\hbar^2)$] of the density and pseudo-spin collective modes, versus energy [in the units of $E_c=\hbar^2/(2 m r_c^2)$] for different values of interlayer distances and at a fixed dimensionless coupling constant $\alpha =15$.\label{fig:p_Dos_rydberg}}
\end{figure}

The density-of-states per unit area (DOS) for each branch of the collective mode reads
\be\label{8}
	\rho_\pm(E ) = \frac{1}{A}\sum_{\bf q} \delta [E  - \hbar{\omega _ \pm }(q)],
\ee
where $A$ is the sample area. 
In Fig.\,\ref{fig:p_Dos_rydberg} we show the density-of-states of symmetric and asymmetric modes for different values of the layer separation. 
At large energies, the free-particle-like dispersion results in a constant DOS, characteristic of the two-dimensional systems. The energies of rotons and maxons are noticeable as singularities in the DOS. 

\section{Results and discussion}\label{sec:results}
In summary, we have investigated the dispersions of collective modes of a double layer structure of bosons with Rydberg-dressed interactions. In the long-wavelength limit, both density and pseudo-spin modes are linear in wave vectors with different sound velocities. At the intermediate values of the wave vector, maxon-roton feature emerges in both of the branches at large coupling constants. 
We have also studied the effects of layer spacing on the dispersion of these modes. At the smaller layer spacings, the asymmetric mode disappears which is characterized by its free-particle-like dispersion. 
Increasing the layer spacing, the two layers are almost decoupled and two modes become degenerate for symmetric bilayers. 
At strong couplings, the homogeneous system becomes unstable towards density waves signaled by the vanishing of the roton energy. Such an instability can appear in either or both of the density and pseudo-spin branches.

A finite superfluid counterflow increases and decreases the energy of the upper and lower excitation branches, respectively. 
Density imbalance between two layers also has a similar effect.
Both counterflow and density imbalance make the homogeneous superfluid phase unstable at smaller coupling constants. 

Recent advances in experimental front~\cite{browaeys2020,de} makes us hopeful that some of the ideas discussed in this work could be also tested in the lab.
 
\acknowledgments
SHA acknowledges the hospitality of the Department of Physics at Bilkent University during the early and final stages of this work. FP and SHA are supported by Iran Science Elites Federation (ISEF). BT acknowledges support from TUBITAK and TUBA. 


\bibliography{Rydberg_bilayer.bbl}

\end{document}